\documentclass[11pt]{article}
\usepackage{xcolor}
\usepackage{textcomp}
\usepackage{cite}
\usepackage{multirow}
\usepackage{enumerate}
\usepackage[scale=0.8]{geometry}

\usepackage{graphicx}
\usepackage[hyperfootnotes=false, linktocpage=true, colorlinks, citecolor=blue, linkcolor=blue, urlcolor=blue, filecolor=blue]{hyperref}
\usepackage[subrefformat=parens,labelformat=parens]{subfig}

\usepackage{amsfonts,amsmath}
\usepackage{multirow,multicol}
\usepackage{bbm}
\usepackage{xspace}
\usepackage{slashed}
\usepackage{parskip}
\newcommand{\bP}{\mathbbm{P}}
\newcommand{\cO}{\mathcal{O}}
\newcommand{\cC}{\mathcal{C}\xspace}
\newcommand{\cZ}{\mathcal{Z}\xspace}

\newcommand{\cA}{\mathcal{A}\xspace}
\newcommand{\cB}{\mathcal{B}\xspace}
\newcommand{\cN}{\mathcal{N}}
\newcommand{\cP}{\mathcal{P}}

\newcommand{\SU}[1]{\ensuremath{\text{SU}(#1)}}
\newcommand{\U}[1]{\ensuremath{\text{U}(#1)}}

\usepackage{hyperref}

\numberwithin{equation}{section}

\begin{document}
\begin{center}
{\Large\bfseries Heterotic Instantons for Monad and Extension Bundles}\\[5mm]

\vspace{1cm}

{\bf Evgeny I. Buchbinder}${}^{\,a,}$\footnote{evgeny.buchbinder@uwa.edu.au},
{\bf Andre Lukas}${}^{\,b,}$\footnote{lukas@physics.ox.ac.uk},
{\bf Burt A. Ovrut}${}^{\,c,}$\footnote{ovrut@elcapitan.hep.upenn.edu},
{\bf Fabian Ruehle}${}^{\,d,b,}$\footnote{fabian.ruehle@physics.ox.ac.uk}

{\small
\vspace*{.5cm}
${}^{a\;}$Department of Physics, The University of Western Australia\\
35 Stirling Highway, Crawley WA 6009, Australia\\ [3mm]
${}^{b\;}$Rudolf Peierls Centre for Theoretical Physics, University of Oxford\\
Parks Road, Oxford OX1 3PU, UK\\[3mm]
${}^{c\;}$Department of Physics and Astronomy, University of Pennsylvania\\ 
Philadelphia, PA 19104-6396, USA\\[3mm]
${}^{d\;}$CERN, Theoretical Physics Department\\
1 Esplanade des Particules, Geneva 23, CH-1211, Switzerland
}
\end{center}
\vspace{1cm}
\begin{abstract}\noindent
We consider non-perturbative superpotentials from world-sheet instantons wrapped on holomorphic genus zero curves in heterotic string theory. 
These superpotential contributions feature prominently in moduli stabilization and large field axion inflation, which makes their presence or absence, 
as well as their functional dependence on moduli, an important issue. We develop geometric methods to compute the instanton superpotentials for heterotic string theory with monad and extension bundles. Using our methods, we find a variety of examples with a non-vanishing superpotential. In view of standard vanishing theorems, we speculate that these results are likely to be attributed to the non-compactness of the instanton moduli space.  We test this proposal, for the case of monad bundles, by considering gauged linear sigma models where compactness of the instanton moduli space can be explicitly checked.  In all such cases, we find that the geometric results are consistent with the vanishing theorems. Surprisingly, linearly dependent Pfaffians even arise for cases with a non-compact instanton moduli space.
This suggests some gauged linear sigma models with a non-compact instanton moduli space may still have a vanishing instanton superpotential.
\end{abstract}

\newpage

\tableofcontents
\setcounter{footnote}{0}

\section{Introduction}
\label{sec:Introduction}
Non-perturbative instanton effects play a crucial role in many aspects of physics, such as moduli stabilization and (large field) axion inflation. 
Especially in light of the de Sitter swampland conjecture~\cite{Obied:2018sgi} and the weak gravity conjecture~\cite{ArkaniHamed:2006dz}, 
it is an important question to study under which conditions these terms can arise and what their moduli dependence is.

In this paper, we will study world-sheet instantons in heterotic theories. These arise from (euclidean) string world-sheets wrapping holomorphic curves of 
genus zero in a Calabi-Yau (CY) manifold $X$. They are dual to D-brane instantons in Type II theories and membrane instantons in M-Theory.\footnote{For recent computations 
of instanton corrections in M-theory see Ref.~\cite{Braun:2018fdp}. For a recent discussion of the prevalence of necessary divisors in toric F-theory setups see~\cite{Halverson:2019vmd}.}

The instanton contributions to the superpotential are, in general, functions of the K\"ahler, the complex structure, and the bundle moduli. Schematically, world-sheet instantons that wrap holomorphic genus zero curves $\gamma_i$, where $i=1,\ldots ,n_\gamma$, with second homology class $\gamma$ contribute a superpotential term
\begin{align}
\label{eq:WSInstantonSuperpotentialContribution1}
W_\gamma=\sum_{i=1}^{n_\gamma} \frac{\text{Pfaff}(\overline{\partial}_{V_{\gamma_i}\otimes\cO_{\gamma_i}(-1)})}{[\text{det}(\overline{\partial}_{\cO_{\gamma_i}})]^2\;\text{det}(\overline{\partial}_{N\gamma_i})}~\text{exp}
\left[ -\int_{\gamma}\frac{J}{2\pi\alpha'}-iB \right]\,.
\end{align}
Let us explain the various terms:
\begin{itemize}
\item $n_\gamma$ is the number of (isolated) holomorphic curves in the curve class of the genus zero curve $\gamma$ as counted by the Gromov-Witten invariants (we will focus on the dominant contribution for which the instanton wraps the curves only once). We denote the $n_\gamma$ curves with class $\gamma$ by $\gamma_i$, where $i=1,\ldots,n_\gamma$.
\item $\text{Pfaff}(\overline{\partial}_{V_{\gamma_i}\otimes\cO_{\gamma_i}(-1)})$ is the Pfaffian of the Dirac operator obtained by integrating over the right-moving fermionic world-sheet degrees of freedom. It depends on the complex structure parameters of the CY and the bundle moduli. Here $V_{\gamma_i}$ denotes the bundle $V$ restricted to $\gamma_i$ and this is tensored with the spin bundle $\cO_{\gamma_i}(-1)$ on $\gamma_i$.
\item $[\text{det}(\overline{\partial}_{\cO_{\gamma_i}})]^2$ is the determinant of the $\overline{\partial}$-operator on the trivial bundle and is just a constant.
\item $\text{det}(\overline{\partial}_{N\gamma_i})$ is the determinant of the $\overline{\partial}$-operator on the normal bundle of $\gamma_i$ which arises from integrating over the bosonic fluctuations on the world-sheet. For a smooth, isolated, genus 0 curve, the normal bundle is just $NC_i=\cO_{\gamma_i}(-1)\otimes\cO_{\gamma_i}(-1)$, 
so this term is $\text{det}(\overline{\partial}_{N\gamma_i})=[\text{det}(\overline{\partial}_{\cO_{\gamma_i}}(-1))]^2$.
\item The (1,1)-forms $J$ and $B$ are the K\"ahler form and the $B$-field on $X$, respectively. The integral over the K\"ahler form 
is the area of the curve, which does not depend on the individual curves $\gamma_i$ but just on their class $\gamma$. The $B$ field cancels the anomalous 
variation of the Pfaffian factor~\cite{Witten:1999eg, Buchbinder:2002pr, Buchbinder:2002ic}.
\end{itemize}

An important observation by Beasley and Witten~\cite{Beasley:2003fx} (see also~\cite{Distler:1992gi, Distler:1993mk, Silverstein:1995re, Basu:2003bq} for related work)
is that, while each instanton contribution associated to a curve $\gamma_i$ can be non-zero, the sum~\eqref{eq:WSInstantonSuperpotentialContribution1} over all instanton contributions in a given class $\gamma$ vanishes, under fairly general conditions. This powerful result is due to a residue theorem in a linear or half-linear $(0,2)$ sigma model. While it would be interesting and, in light of the swampland conjectures, important to see how this condition carries over to other string theories, such as Type II under heterotic-Type II duality, we will focus on the heterotic case.

The residue theorem of Beasley--Witten is based on the following assumptions. The compactification Calabi--Yau (CY) manifold $X$, its associated K\"ahler form, $J$ and 
the vector bundle on $X$ must descend from a projective or, more generally, toric ambient space. Additionally, the instanton moduli space must be ``compact''.  This condition is defined in terms of the linear or half-linear $(0,2)$ sigma model and manifests itself in a possible appearance of additional fermionic zero modes which lead to the vanishing of the instanton sum~\eqref{eq:WSInstantonSuperpotentialContribution1}. Within this context of linear and half-linear sigma models, it is not  easily checked.

Bertolini and Plesser~\cite{Bertolini:2014dna} have emphasized the importance of compactness for the validity of the residue theorem. In cases with a gauged linear sigma model (GLSM)~\cite{Witten:1993yc} description, they have formulated a criterion that allows checking compactness of the instanton moduli space.
More specifically, their compactness criterion is simply a condition on the GLSM \U1 charges. If the instanton moduli space turns out to be non-compact, the residue theorem of Beasley and Witten does not apply and one cannot conclude that the instanton sum vanishes. 

To the best of our knowledge, it is not known whether every heterotic compactification can be described as a GLSM. In a GLSM, the vector bundle $V$ is naturally described in terms of a monad bundle. Besides this bundle construction, there exist other descriptions
such as spectral cover or extension bundles, and it is unclear whether every such bundle can re-written as a monad bundle. Moreover, a generic heterotic compactification involves five-branes and it is not clear how they can be incorporated into a GLSM description (see however~\cite{Blaszczyk:2011ib,Quigley:2011pv}). Consequently, checking Bertolini and Plesser's compactness criterion can still be a difficult task for models which are not originally formulated in terms of a GLSM.

In Refs.~\cite{Braun:2008sf,Braun:2007vy,Braun:2007xh,Braun:2007tp,Buchbinder:2016rmw,Buchbinder:2017azb, Buchbinder:2018hns}, tools were derived that facilitate the computation of Pfaffians purely based on methods of algebraic geometry. The key observation is that the Pfaffian on some curve $\gamma_i$ will be zero iff the bundle
\begin{align}
\label{eq:DefinitionVi}
V_i:=V|_{C_i}\otimes\cO_{C_i}(-1)
\end{align}
has global sections. Hence, instead of computing the Pfaffian directly, we can compute the cohomology dimension $h^0(V_i)$. This dimension depends on both complex structure and bundle moduli. Here, we fix the complex structure moduli to a suitable generic value and study the dependence on the bundle moduli~$\beta$. If $h^0(V_i)>0$ everywhere in bundle moduli space, the Pfaffian vanishes identically. A more interesting case arises when $h^0(V_i)=0$ generically, but if there exists a ``jumping locus" in bundle moduli space, typically described by an equation $\delta_i(\beta)=0$ for a polynomial $\delta_i$, for which $h^0(V_i)>0$. Since $\delta_i$ and the Pfaffian have the same zero locus they must be proportional, so that
\begin{align}
\delta_i= \lambda_i\text{Pfaff}(\bar\partial_{V_i})\; ,
\label{z1}
\end{align}
for $\lambda_i\in\mathbbm{C}^*$. The superpotential term~\eqref{eq:WSInstantonSuperpotentialContribution1}, we can then be written as
\begin{align}
\label{eq:WSInstantonSuperpotentialContribution2}
W_\gamma= \left[\sum_{i=1}^{n_\gamma}\lambda_i \delta_i\right]\exp\left(-\int_\gamma \frac{J}{2\pi\alpha'}+iB\right)\; .
\end{align}
We can use this result to formulate a sufficient condition for the non-vanishing of the instanton contribution to the superpotential. If all $\delta_i$ are 
linearly independent as functions of the bundle moduli $\beta$, then the instanton superpotential is necessarily non-vanishing. 

In the present paper, we apply this criterion to study the vanishing/non-vanishing of instanton superpotentials in the geometric framework. We find, perhaps surprisingly, that compactifications with non-vanishing superpotential are more common than the apparent generality of the residue theorem seems to suggest. We do not know the precise reason for why the residue theorem does not apply to those cases, but we propose that it is due to a violation of the compactness assumption. In order to support this proposal, we study the relation between geometric models with monad bundles and the associated GLSM formulation.

The remainder of the paper is organized as follows. In Section~\ref{sec:GeometricApproach}, we will recapitulate the geometric setting, including the construction of the CY manifolds, various bundle constructions, the method to compute the curves $\gamma_i$ and the geometric method to compute Pfaffians outlined above.
We will also review a technique~\cite{Buchbinder:2013dna, Buchbinder:2014qda} that allows us to find a monad descriptions of certain 
extension bundles which will be used later. In Section~\ref{sec:Examples} we present geometric examples of heterotic compactifications in which all geometric data 
descends from a projective ambient space, yet the instanton sum~\eqref{eq:WSInstantonSuperpotentialContribution1} does not vanish.
The non-cancellation simply follows from linear independence of the holomorphic functions $\delta_i$ 
in Eqs.~\eqref{z1}, \eqref{eq:WSInstantonSuperpotentialContribution2}. We present several examples both for extension and monad bundles to illustrate this. In a companion paper~\cite{Buchbinder:2019aaa} we report on a systematic scan over hundreds of thousands of models which shows that such examples are quite common. 
In Section~\ref{sec:GLSMs}, we review the GLSM description of heterotic models with an emphasis on the Bertolini--Plesser criterion for compactness of
the instanton moduli space. In Section~\ref{sec:Compactness} we discuss geometric models for which we can find a GLSM description and we compare the statements about instanton superpotentials which can be extracted in either framework. We conclude in Section~\ref{sec:Conclusions}.

\section{Instanton superpotentials from geometry}
\label{sec:GeometricApproach}
\subsection{The Calabi-Yau geometry}
\label{sec:CY}
In Ref.~\cite{Buchbinder:2017azb}, we have described a method of identifying all genus zero curves within certain homology classes of CICY manifolds. This method applies if the homology class under consideration descends from a $\mathbbm{P}^1$ factor of the ambient space. In this case, the union of all genus zero curves in this class can be written as a complete intersection. We will now briefly review this method.

Our manifolds are defined in an ambient space $\mathcal{P}$ which we take to be a product
\begin{align}
\mathcal{P}=\mathbbm{P}^{n_1}\times\ldots\times\mathbbm{P}^{n_m}\,.
\label{e1}
\end{align}
of $m$ projective factors. Within such an ambient space, a family of CICYs, $X$, is defined by a configuration matrix
\begin{align}
\label{e10}
X\in\left[
\begin{array}{c|cccc}
\mathbbm{P}^{n_1}&q_1^1&q_2^1&\ldots&q_K^1\\
\mathbbm{P}^{n_2}&q_1^2&q_2^2&\ldots&q_K^2\\
\vdots&\vdots&\vdots&\ddots&\vdots\\
\mathbbm{P}^{n_m}&q_1^m&q_2^m&\ldots&q_K^m
\end{array}
\right]\; ,
\end{align}
which specifies the multi-degrees of the defining polynomials. More specifically, the CICY is defined as the common zero locus of $K$ polynomials $p_a$, where the multi-degree of $p_a$ is given by the $a^{\rm th}$ column of the configuration matrix, so by $q_a=(q_a^1,q_a^2,\ldots ,q_a^m)^T$. The coefficients in generic polynomials $p_a$ of the appropriate degree parametrize the complex structure of $X$. Not all these coefficients affect the complex structure (for example the overall scaling of the coefficients in $p_a$ is irrelevant) but this redundancy can be removed by suitably fixing some of the coefficients.

Since we are interested in CY three-folds we require that $K+3=\sum_{i=1}^m n_i$. The CY condition, $c_1(TX)=0$, amounts to
\begin{align}
\label{eq:CICY_CY_Condition}
\sum_{a=1}^K q^i_a \stackrel!= n_i+1\,,\qquad i=1,2,\ldots, m\,.
\end{align}
For our discussion, we assume that the ambient space $\mathcal{P}$ contains at least one $\mathbb{P}^1$ factor  which corresponds to the homology class $\gamma$ we would like to study. We also order the projective factors such that this $\mathbb{P}^1$ appears first and write the ambient space as $\mathcal{P}=\mathbb{P}^1\times\mathcal{Q}$, where $\mathcal{Q}=\mathbb{P}^{n_2}\times\cdots\times\mathbb{P}^{n_m}$. Further, it is convenient to denote the degrees related to the ``transverse space" $\mathcal{Q}$ by $\hat{q}_a=(q_a^2,\ldots, q_a^m)^T$.

Due to the CY condition~\eqref{eq:CICY_CY_Condition}, there are (up to trivial reordering) only two possibilities for the degrees in a $\mathbb{P}^1$ direction. This leads to two types of configuration matrices~\cite{Buchbinder:2017azb}, referred to as type I and type II, given by
\begin{align}
\label{eq:CurvesInBothCases}
\text{type I}:\; X\in\left[
\begin{array}{c|cc	ccc}
\mathbbm{P}^{1}&1&1&0&\ldots&0\\
\mathcal{Q}&\hat{q}_1&\hat{q}_2&\hat{q}_3&\ldots&\hat{q}_K\\
\end{array}
\right]\,,\qquad
\text{type II}:\; X\in\left[
\begin{array}{c|ccc	c}
\mathbbm{P}^{1}&2&0&\ldots&0\\
\mathcal{Q}&\hat{q}_1&\hat{q}_2&\ldots&\hat{q}_K
\end{array}
\right]\,.
\end{align}
Associated to these two types, we can define the complete intersection
\begin{align}
\label{eq:CICYCurves}
\begin{split}
\text{type I}:&~~ \{y_I\}\in\left[
\begin{array}{c|cc	ccccc}
\mathcal{Q}&\hat{q}_1&\hat{q}_1&\hat{q}_2&\hat{q}_2&\hat{q}_3&\ldots&\hat{q}_K
\end{array}
\right]\,,\\
\text{type II}:&~~ \{y_I\}\in\left[
\begin{array}{c|cccccc}
\mathcal{Q}&\hat{q}_1&\hat{q}_1&\hat{q}_1&\hat{q}_2&\ldots&\hat{q}_K\\
\end{array}
\right]\,.
\end{split}
\end{align}
in the transverse space $\mathcal{Q}$. By counting dimension, it is easy to check that these complete intersections are zero-dimensional and, hence, correspond to a finite number of points $y_i\in\mathcal{Q}$, where $i=1,\ldots ,n_\gamma$. Provided they are based on the polynomials which descend from the original CY configurations~\eqref{eq:CurvesInBothCases}, it can be verified that the $n_\gamma$ genus zero curves $\gamma_i=\mathbb{P}^1\times y_i\subset X$ are contained in the CY manifold $X$ and have homology class $\gamma$. Moreover, comparison with the Gromov-Witten invariants for $\gamma$ shows that these provide all the genus zero curves in this class.

In summary, we  have an explicit method to find all genus zero curves in homology classes $\mathcal{C}$ associated to ambient space $\mathbb{P}^1$ factors. Specifically, the locations $y_I$ of these curves in the transverse space $\mathcal{Q}$ are determined by the configuration matrices~\eqref{eq:CICYCurves}.


\subsection{The bundles}
\label{sec:bundles}
For the construction of geometric heterotic vacua we require vector bundles $V\rightarrow X$ with structure groups that can be embedded into $E_8$. In particular, this means that $c_1(V)=0$. We also require that $c_2(TX)-c_2(V)$ is in the Mori cone of $X$ so that there is a guaranteed solution to the  heterotic anomaly condition in terms of five-branes (although adding a ``hidden" bundle might be possible as well). For $V$ to be supersymmetric it needs to be poly-stable, a condition which can be checked algorithmically~\cite{CICYPackage}, although the process can in practice be very tedious.

Throughout the paper, we will use different bundle constructions but our basic building blocks will always be line bundles on the ambient space $\mathcal{P}$, which we denote by ${\cal O}_{\mathcal{P}}(k)$ with multi-degree $k=(k^1,\ldots ,k^m)$, and their restrictions ${\cal O}_X(k)={\cal O}_{\mathcal{P}}(k)|_X$ to the CY manifold $X$. Note that the defining equations $p_a$ of the CY manifold $X$ are sections of the bundle $\cN=\bigoplus_a\cN_a$, where $\cN_a={\cal O}_\cP(q_a)$. If the entire second cohomology of $X$ descends from the second ambient space cohomology, the CICY is called {\em favorable}. In this case, all line bundles on $X$ are obtained by restricting the line bundles on $\mathcal{P}$.

We will frequently require line bundle sums which we denote by
\begin{equation}
 {\cal A}=\bigoplus_{\alpha=1}^{r_A}{\cal O}_\mathcal{P}(a_\alpha)\;,\qquad
 {\cal B}=\bigoplus_{\beta=1}^{r_B}{\cal O}_\mathcal{P}(b_\beta)\;,\qquad
 {\cal C}=\bigoplus_{\gamma=1}^{r_C}{\cal O}_\mathcal{P}(c_\gamma)\; ,
\end{equation}
with multi-degrees $a_\alpha=(a_\alpha^1,\ldots ,a_\alpha^m)$,  $b_\beta=(b_\beta^1,\ldots ,b_\beta^m)$ and $c_\gamma=(c_\gamma^1,\ldots ,c_\gamma^m)$. Their restrictions to the CY manifold are denoted by the corresponding non calligraphic letter, so
\begin{equation}
A={\cal A}|_X=\bigoplus_{\alpha=1}^{r_A}{\cal O}_X(a_\alpha)\;,\qquad
B={\cal B}|_X=\bigoplus_{\beta=1}^{r_B}{\cal O}_X(b_\beta)\\;,\qquad
C={\cal C}|_X=\bigoplus_{\gamma=1}^{r_C}{\cal O}_X(c_\gamma)\; . \label{lbsX}
\end{equation}
The first and second Chern character of a line bundle sum can be computed from
\begin{equation}
 {\rm ch}_1^i(A)=\sum_{\alpha=1}^{r_A}a_\alpha^i\;,\qquad{\rm ch}_{2i}(A)=\frac{1}{2}d_{ijk}\sum_{\alpha=1}^{r_A}a_\alpha^j a_\alpha^k\; ,
\end{equation}
and similarly for $B$ and $C$. Here, $d_{ijk}$ are the triple intersection numbers of $X$.

In the following, we will usually write down the defining sequences for the bundle $V$ on $X$ but it is understood that associated sequences exist on the ambient space, defining a bundle (or, in some cases a sheaf) $\mathcal{V}$ that restricts to $V={\cal V}|_X$. 

The tangent space to the bundle moduli space ${\cal M}_X(V)$ of $V$ is computed by $H^0(V\otimes V^*)$. If all bundle moduli of $V$ descend from moduli of $\mathcal{V}$ we call the bundle {\em favorable}, in analogy with the corresponding terminology for the manifold. For our calculations we will require moduli to descend from the ambient space, so for favorable bundles we will be able to deal with the entire moduli space. For non-favorable bundles our results will be restricted to the subset of ${\cal M}_X(V)$ which does descend from $\mathcal{V}$.

We begin the more explicit discussion with extension bundles. 


\subsubsection*{Simple extension bundles}
The simplest type of extension bundle we will consider is an extension of two line bundle sums, defined by the short exact sequence
\begin{equation}
 0\to B\to V\to C\to 0\; , \label{extseq}
\end{equation} 
and with Chern character
\begin{eqnarray}
 {\rm rk}(V)&=&r_B+r_C\\
 {\rm ch}_1^i(V)&=&{\rm ch}_1^i(B)+{\rm ch}_1^i(C)=\sum_\beta b_\beta^i+\sum_\gamma c_\gamma^i\stackrel{!}{=}0\\
 {\rm ch}_{2i}(V)&=&{\rm ch}_{2i}(B)+{\rm ch}_{2i}(C)=\frac{1}{2}d_{ijk}\left[\sum_\beta b_\beta^j b_\beta^k+\sum_\gamma c_\gamma^j c_\gamma^k\right]\; ,
\end{eqnarray} 
where the last equation assumes that ${\rm ch}_1(V)=0$. The moduli space of such extension bundles is
\begin{equation}
 \mathcal{M}_X(V)={\rm Ext}^1(C,B)\cong H^1(C^*\otimes B)\; .
\end{equation} 
The zero in ${\rm Ext}^1(C,B)$ corresponds to the trivial extension $V=B\oplus C$ and non-trivial extensions (with non-Abelian structure groups) are possible if ${\rm Ext}^1(C,B)$ is non-trivial. 

For instanton calculations we need an explicit handle on this bundle moduli space. The most straightforward case is the one where $ \mathcal{M}_X(V)$ happens to be equal to its ambient space counterpart $ \mathcal{M}_\mathcal{P}(\mathcal{V})={\rm Ext}^1(\mathcal{C},\mathcal{B})$. Things are not always this simple, however. There are two effects which can lead to a difference between the moduli spaces $\mathcal{M}_X(V)$ and $\mathcal{M}_\mathcal{P}(\mathcal{V})$. 

A bundle can be non-favorable if $\mathcal{M}_X(V)$ receives contributions from cohomologies other than $\mathcal{M}_\mathcal{P}(\mathcal{V})\cong H^1(\mathcal{C}^*\otimes\mathcal{B})$. In practice, we can only handle bundle moduli which descend from ambient space moduli, so for such non-favorable bundles we are only be able to consider the sub-space of $\mathcal{M}_X(V)$ which descends. 

On the other hand, considering the ambient moduli space $\mathcal{M}_\mathcal{P}(\mathcal{V})$ might also be over-counting, since restriction to $X$ can imply that certain quotients have to be formed to obtain the correct moduli space $\mathcal{M}_X(V)$. This can be corrected for relatively easily by identifying and carrying out the relevant quotients which arise in the Koszul sequence. 

We will ensure supersymmetry of the extension bundle in a neighborhood of the trivial extension by ensuring that the line bundle sum $B\oplus C$  is superymmetric. This amounts to checking that there is a common solution to the slope zero condition
\begin{equation}
 \mu_X(L)=\int_X J^2\wedge c_1(L)\stackrel{!}{=}0 \label{slope0}
\end{equation}
(where $J$ is the K\"ahler form of $X$) for all line bundles $L\subset B\oplus C$.

\subsubsection*{Double extensions}
On occasion, we will consider more complicated extension bundles $V$ which are defined by two exact sequences
\begin{equation}
 0\to A\to V'\to B\to 0\;,\qquad 0\to V'\to V\to C\to 0\; , \label{doubleext}
\end{equation} 
which lead to the Chern character
\begin{eqnarray}
 {\rm rk}(V)&=&r_A+r_B+r_C\\
 {\rm ch}_1^i(V)&=&{\rm ch}_1^i*(A)+{\rm ch}_1^i(B)+{\rm ch}_1^i(C)=\sum_\alpha a_\alpha^i+\sum_\beta b_\beta^i+\sum_\gamma c_\gamma^i\stackrel{!}{=}0\\
 {\rm ch}_{2i}(V)&=&{\rm ch}_{2i}(A)+{\rm ch}_{2i}(B)+{\rm ch}_{2i}(C)=\frac{1}{2}d_{ijk}\left[\sum_\alpha a_\alpha^j a_\alpha^k +\sum_\beta b_\beta^j b_\beta^k+\sum_\gamma c_\gamma^j c_\gamma^k\right]
\end{eqnarray} 
As before, the last equation is valid provided ${\rm ch}_1(V)=0$.

There are two bundle moduli spaces involved, namely
\begin{equation}
 \mathcal{M}_X(V)={\rm Ext}^1(C,V')\cong H^1(C^*\otimes V')\; ,\qquad
 \mathcal{M}_X(V')={\rm Ext}^1(B, A)\cong H^1(B^*\otimes A) \label{ext2}
\end{equation}
and choosing the zero in either moduli space leads to the trivial extension $V=A\oplus B\oplus C$. If the extension ${\rm Ext}^1(B,A)=0$ then $V'=A\oplus B$ and the double extension reduces to the single extension $0\to A\oplus B\to V\to C\to 0$. Only if both extension groups in Eq.~\eqref{ext2} are non-zero can we have a non-trivial double extension where both $V'$ and $V$ are bundles with non-Abelian structure group. 

The moduli space $ \mathcal{M}_X(V)$ can be computed from the short exact sequence
\begin{equation}
 0\to C^*\otimes A\to C^*\otimes V'\to C^*\otimes B\to 0\; ,
\end{equation} 
obtained by tensoring the first sequence~\eqref{doubleext} by $C^*$, and its associated long exact sequence in cohomology. Explicit expressions depend somewhat on the circumstances. For example, if $H^0(C^*\otimes B)=H^2(C^*\otimes A)=0$ then
\begin{equation}
  \mathcal{M}_X(V)\cong H^1(C^*\otimes V')\cong H^1(C^*\otimes A)\oplus H^1(C^*\otimes B)\; .
\end{equation}  

An explicit description of the moduli space in terms of the moduli space of the ambient space bundle has the same issues as discussed for the single extension case. For non-favorable double extensions bundles we can only include the part of the moduli space $ \mathcal{M}_X(V)$ in our instanton calculation which descends from $\mathcal{M}_\mathcal{P}(\mathcal{V})$. Over-counting which arises from quotients in the Koszul sequence can be taken into account explicitly. 

Just as in the case of a single extension, supersymmetry of bundles $V$ obtained from a double extension is ensured, in a neighborhood of the trivial extension, by ensuring a common solution to Eq.~\eqref{slope0} for all line bundles $L\subset A\oplus B\oplus C$.

\subsubsection*{Monad bundles}

We will focus on two-term monads and define the monad vector bundle $V$ by the short exact sequence
\begin{equation}
\label{eq:TwoTermMonad}
0\to V\to A\xrightarrow{f} B\to 0\quad\Rightarrow\quad V={\rm Ker}(f)\; .
\end{equation}
where $A$ and $B$ are line bundles sums as in Eq.~\eqref{lbsX}. The monad map $f$ can be thought of as an $r_B\times r_A$ matrix whose entries are sections 
\begin{equation}
f_{\beta\alpha}\in H^0({\cal O}_X(b_\beta-a_\alpha))\; . \label{fab}
\end{equation}
In practice, we will only incorporate those sections in $f_{\beta\alpha}$ which descend from sections $H^0({\cal O}_\mathcal{P}(b_\beta-a_\alpha))$, that is, which can be written down as polynomials in the ambient space coordinates. For $V$ to be a bundle, rather than a sheaf, we need to require that the monad map $f$ does not degenerate anywhere on $X$. The Chern character of a monad bundle is given by
\begin{eqnarray}
 {\rm rk}(V)&=&r_A-a_B\\
 {\rm ch}_1^i(V)&=&{\rm ch}_1^i(A)-{\rm ch}_1^i(B)=\sum_\alpha a_\alpha^i-\sum_\beta b_\beta^i\stackrel{!}{=}0\\
 {\rm ch}_{2i}(V)&=&{\rm ch}_{2i}(A)-{\rm ch}_{2i}(B)=\frac{1}{2}d_{ijk}\left[\sum_\alpha a_\alpha^j a_\alpha^k-\sum_\beta b_\beta^j b_\beta^k\right]\; ,
\end{eqnarray} 
where the last equation holds provided ${\rm ch}_1(V)=0$. 

The expression for the dimension $h^0(V\otimes V^*)$ of the bundle moduli space is in general complicated. However, provided we have
\begin{equation}
h^1(B^*\otimes B)=0\,,\qquad h^1(B^*\otimes A)=0\; , \label{VVcond}
\end{equation}
which is frequently satisfied for concrete models, there is a simple formula~\cite{Anderson:2008uw}:
\begin{align}
\label{eq:MonadBundleEndDim}
\begin{split}
h^0(V^*\otimes V)&=\phantom{-\;}h^0(A^*\otimes B)+h^0(B^*\otimes A)-h^0(A^*\otimes A)-h^0(B^*\otimes B)\\
&\phantom{=\;}-h^1(B^*\otimes A) + h^1(A^*\otimes A) + 1\,.
\end{split}
\end{align}
The terms in this formula which involve $h^1(\cdot)$, as well as $h^0(B^*\otimes A)$, correspond to non-polynomial deformations. If they are non-vanishing the bundle is non-favorable. The remaining terms have a straightforward interpretation: $h^0(A^*\otimes B)$ counts the number of sections~\eqref{fab} which enter the monad map. For the bundle to be favorable all these sections have to descend from sections on the ambient space, so be polynomial in the ambient space coordinates. The two terms $h^0(A^*\otimes A)$ and $h^0(B^*\otimes B)$ count the number of bundle automorphisms of $A$ and $B$ which have to subtracted for a correct moduli count. Finally, $+1$ is added to correct for the overall scaling which has been subtracted twice. 

\subsubsection*{Relation between monad and extension bundles}
\label{sec:ExensionMonadMap}
A given bundle construction, even if favorable, does not necessarily exhaust the entire moduli space of a given topological bundle type.
For example, an extension and a monad bundle might realize the same topological type but they can correspond to different parts of the bundle moduli space. 
For this reason, it can be useful to convert between the different constructions. Here, we describe how to find a monad description of an extension bundle, 
following the method of Ref.~\cite{Buchbinder:2013dna, Buchbinder:2014qda}. 

The basic idea is to start with a monad description of the structure sheaf,
\begin{align}
\label{eq:MonadStartingPoint}
0\rightarrow \cO_X\rightarrow\tilde{A}\xrightarrow{\phi}\tilde{B}\rightarrow 0\,,
\end{align}
where $\tilde{A}$, $\tilde{B}$ are line bundle sums such that
\begin{align}
\label{eq:TrivialBundle}
\text{rk}(\tilde{A})=\text{rk}(\tilde{B})+1 \quad\text{~and~}\quad c_1(\tilde{A})=c_1(\tilde{B})\,.
\end{align}
These conditions are necessary for the monad to describe the trivial bundle but not sufficient. We also need to ensure that the monad map $\phi$ does not degenerate on $X$, so that its kernel is indeed a bundle rather than a sheaf. To illustrate how this can be done, we consider our preferred ambient spaces of the form $\mathcal{P}=\mathbb{P}^1\times\mathcal{Q}$. For such a case, we choose for the above line bundles sums
\begin{align}
\tilde{A}=\cO_X(r,0,\ldots,0)\oplus\cO_X(\tilde{r},0,\ldots,0)\,,\qquad \tilde{B}=\cO_X(r+\tilde{r},0,\ldots,0)\,, \label{AtBt}
\end{align}
where $r$ and $\tilde{r}$ are positive integers. Then, the monad map is of the form $\phi=(\varphi,\tilde{\varphi})$, where $\varphi$ and $\tilde{\varphi}$ are polynomials of degree $\tilde{r}$ and $r$, respectively, in the coordinates of the $\mathbb{P}^1$. Provided $\varphi$ and $\tilde{\varphi}$ are sufficiently generic they have no common zero locus in $\mathbb{P}^1$ so the map does indeed not degenerate. 

Having obtained a monad for the trivial bundle it is then straightforward to write down a monad for a line bundle sum $V=\bigoplus_aL_a$ as
\begin{equation}
 0\to V\to \bigoplus_a L_a\otimes\tilde{A}_a\stackrel{F}{\to}\bigoplus_aL_a\otimes \tilde{B}_a\to 0\; , \label{extmon}
\end{equation}
where $\tilde{A}_a$ and $\tilde{B}_a$ are line bundle sums of the form~\eqref{AtBt}, but possibly with different values of $r$ and $\tilde{r}$ for different $a$. To represent this line bundle sum the monad map $F$ should have the structure
\begin{equation}
 F=\left(\begin{array}{ccccc}\varphi_1&\tilde{\varphi}_1&0&0&\cdots\\0&0&\varphi_2&\tilde{\varphi}_2&\cdots\\\vdots&\vdots&\vdots&\vdots&\vdots\end{array}\right) \label{Fmap}
\end{equation} 
where $\phi_a=(\varphi_a,\tilde{\varphi}_a)$ are the maps in the monad realization~\eqref{eq:MonadStartingPoint} of the trivial bundle (with $\tilde{A}$ and $\tilde{B}$ replaced by $\tilde{A}_a$ and $\tilde{B}_a$).

Now suppose we have an extension~\eqref{extseq} of two line bundle sums $B$ and $C$. Following the above prescription we can then find a monad description for the trivial extension $V=B\oplus C$. The point is that the monad map in this description can often be deformed away from the simple pattern in Eq.~\eqref{Fmap} by filling in some of the zero entries. In this way, the monad can be deformed away from the line bundle locus.
 
\subsection{Computing the Pfaffians}
Let us finally outline the general procedure to compute the Pfaffians. Details, such as finding a parametrization of the bundle moduli space or computing relevant   cohomologies, are model-dependent and will be illustrated by the examples in Section~\ref{sec:Examples} and Section~\ref{sec:Compactness}.

Recall from Eq.~\eqref{eq:DefinitionVi} the definition of the bundles $V_i=V|_{\gamma_i}\otimes{\cal O}_{\mathbb{P}^1}(-1)$, where $\gamma_i$ is a holomorphic genus zero curve.
The individual Pfaffians for curves $\gamma_i$ vanish if the operator $\bar\partial_{V_i}$ has zero modes. Since the zero modes of this operator are counted by the sections of the line bundle $V_i$, we get
\begin{align}
h^0(V_i)\neq 0\qquad\Leftrightarrow\qquad \text{Pfaff}(\bar\partial_{V_i})= 0\,.
\end{align}
Note that the cohomologies $H^0(V_i)$ depend on both bundle and complex structure moduli. To simplify the computation, we will fix the complex structure at a sufficiently generic value and study the dependence of $H^0(V_i)$ on the bundle moduli $\beta$. If $h^0(V_i)>0$ everywhere in bundle moduli space then the Pfaffian for $\gamma_i$ vanishes identically. A more interesting case arises when $h^0(V_i)=0$ for generic values but there is a ``jumping locus", a locus of complex co-dimension one in bundle moduli space, where $h^0(V_i)>0$. Such a jumping locus can be described as the zero locus of a polynomial $\delta_i=\delta_i(\beta)$ and we conclude that $\text{Pfaff}(\bar\partial_{V_i})=\lambda_i\delta_i$ for a complex constant $\lambda_i$. The procedure to compute instanton superpotential geometrically can, hence, be summarized as follows.
\begin{enumerate}
\item Pick a curve class $\gamma$ in the second homology of $X$ for which the contribution $W_\gamma$ to the instanton superpotential is to be computed.
\item Find a set of holomorphic representatives of all genus zero curves $\gamma_i\subset X$, where $i=1,\ldots ,n_\gamma$, in the curve class $\gamma$.
\item Restrict the vector bundle $V\rightarrow X$ to each curve in the given class and tensor it with $\mathcal{O}_{\mathbb{P}^1}(-1)$, that is, compute the bundles $V_i=V|_{\gamma_i}\otimes{\cal O}_{\mathbb{P}^1}(-1)$.
\item If $h^0(V_i)>0$ everywhere in bundle moduli space, then $\text{Pfaff}(\bar\partial_{V_i})= 0$.
\item Otherwise, if  $h^0(V_i)=0$ generically, find the jumping locus where $h^0(V_i)>0$ and describe it as the zero locus of a polynomial $\delta_i=\delta_i(\beta)$.  Then $\text{Pfaff}(\bar\partial_{V_i})=\lambda_i \delta_i$ for some complex constant $\lambda_i$.
\item The instanton superpotential contribution for the class $\gamma$ is then given by
\begin{equation}
 W_\gamma=\left[\sum_{i=1}^{n_\gamma}\lambda_i\delta_i\right]\exp\left(-\int_\gamma(J+iB)\right)\; . \label{WC}
\end{equation} 
\item Check if the polynomials $\delta_i$ are linearly independent, as functions of the bundle moduli $\beta$. If they are, the pre-factor in Eq.~\eqref{WC} cannot vanish identically and $W_\gamma$ is non-zero. If they are linearly dependent, we cannot make a definite statement. In this case, $W_\gamma$  can be zero or non-zero, depending on the unknown constants $\lambda_i$.
\end{enumerate}


\section{Non-vanishing superpotentials in the geometric approach}
\label{sec:Examples}


The main purpose of this section is to present a number of explicit examples which lead to a non-vanishing instanton superpotential and, at the same time satisfy the following conditions which underlie the Beasley--Witten residue theorem:
\begin{itemize}
\item The Calabi--Yau manifold $X$ is a CICY in some projective ambient space $\mathcal{P}$, as in Eq.~\eqref{e1}. 
\item The vector bundle on $X$ is a restriction of a vector bundle on $\mathcal{P}$. 
\item The K\"ahler form on $X$ is a restriction of a K\"ahler form on $\mathcal{P}$.\footnote{This assumption is not mentioned in Ref.~\cite{Beasley:2003fx}, 
and was established later, in Ref.~\cite{Buchbinder:2016rmw}.}
\end{itemize}
In a companion paper~\cite{Buchbinder:2019aaa}, we are reporting the result of systematic computer scans of hundreds of thousands of heterotic models on more than 100 different CICYs, mostly with $h^{1, 1}=2, 3$.  They lead to many thousands of models which satisfy the above conditions and where the contributions from individual instantons are linearly independent and thus cannot cancel.  Here, we will illustrate this with a number of explicit examples.

At the end of this section, we will speculate about possible reasons why the examples in this section -- and indeed the large classes presented in Ref.~\cite{Buchbinder:2019aaa} -- avoid the residue theorem.

Some technical details of our examples, such as the precise location of the holomorphic genus zero curves and equations of  the sub-loci $\delta_i=0$ which  
determine the zeroes of the Pfaffians, are given by very cumbersome expressions.  These expressions are  computed with Mathematica and will not be presented explicitly.


\subsection{An example with an SU(3) monad bundle}
\label{sec:Monad_Example}


We will start with the following example. We consider CICY 7735 in the list of Ref.~\cite{Candelas:1987kf} which is described by the configuration matrix
\begin{align}
\label{eq:CICY7735Conf}
X\in\left[
\begin{array}{c|cccc}
 \mathbbm{P}^1& 2 & 0 & 0 & 0 \\
 \mathbbm{P}^1& 0 & 2 & 0 & 0 \\
 \mathbbm{P}^5& 1 & 1 & 2 & 2 \\
\end{array}
\right]^{(3,51)}
\qquad
\left|
\qquad
\begin{array}{l}
z_{1,0}, z_{1,1}\\
z_{2,0}, z_{2,1}\\
z_{3,0}, z_{3,1},z_{3,2}, z_{3,3},z_{3,4}, z_{3,5}~\ .
\end{array}
\right.
\end{align}
We find that the genus zero Gromov-Witten invariants for the three divisors are~$\vec{n}_{\gamma}=(8,8,128)$. The geometry is of Type II as 
discussed in Section~\ref{sec:CY}, and we can use the method described there to find the position of the eight curves in the class associated to the first $\mathbbm{P}^1$  factor. In particular, these eight points are described by the complete intersection
\begin{equation}
\{y_{i}\}= \left[\begin{array}{c|cccccc}\mathbb{P}^1&0&0&0&2&0&0\\\mathbb{P}^5&1&1&1&1&2&2\end{array}\right]\;,
\end{equation}
with $y=(\vec{z}_2, \vec{z}_{3})$.

\subsubsection*{The bundle}
The monad bundle $V$ on $X$ we consider is defined by the short exact sequence $0\to V\to A\stackrel{f}{\to} B\to 0$ with
\begin{align}
A=\mathcal{O}_X(1,1,0)\oplus\mathcal{O}_X(1,1,1)\oplus\mathcal{O}_X(0,1,0)\oplus\mathcal{O}_X(0,1,0)\;,\qquad B=\mathcal{O}_X(2,4,1)\; .
\end{align}
This bundle has the following properties:
\begin{itemize}
  \item It satisfies $c_1(V)=0$ and $c_2 (TX) - c_2 (V)$ is effective.
  \item It satisfies basic stability checks.
  \item For sufficiently generic choices of the monad map it is a vector bundle rather than a sheaf.
\end{itemize}

The monad map $f$ can be written as a $1\times4$ matrix
\begin{align}
\label{eq:monadMapFull}
 f=\left(
 f_{1,3,1}\,,\; f_{1,3,0}\,,\; f_{2,3,1}\,,\; f'_{2,3,1}\right)\,,
\end{align}
where the subscripts denote the multi-degrees of the polynomial maps. Since the third and fourth function have the same multi-degrees, but correspond to independent maps, we denote them by $ f_{2,3,1}$ and $f'_{2,3,1}$, respectively.

\subsubsection*{The bundle moduli space and redundancies}
Let us next study the monad bundle moduli space following the procedure outlined in Section~\ref{sec:bundles}. We first compute $h^1(X,B^*\otimes B)$ and $h^1(X,B^*\otimes A)$ to check that Eq.~\eqref{VVcond} is satisfied, which is indeed the case. From Eq.~\eqref{eq:MonadBundleEndDim} we then find $h^{1}(X,V\otimes V^*)=130$. This should be compared with the number of parameters in the polynomial map~\eqref{eq:monadMapFull}. We find that $f_{1,3,1}$ has 48 monomials,  $f_{1,3,0}$ has 8, and $f_{2,3,1}$ and $f'_{2,3,1}$ have 72 each, giving a total of 200 parameters. Thus this example involves an over-parametrization of  the bundle moduli space.

To study this in more detail, we compute the individual terms in Eq.~\eqref{eq:MonadBundleEndDim}. We first note that some of the monomials in the ambient space monad map restrict trivially to $X$. This is reflected in the fact that
\begin{align}
h^0(\cO_X(1,3,1))=44\,,\qquad h^0(\cO_X(1,3,0))=8\,,\qquad h^0(\cO_X(2,3,1))=62\,.
\end{align}
So, for entry in the monad map, only 44 of the 48 ambient space monomials restrict non-trivially. This can be seen from the Koszul resolution. Indeed, we find that 
\begin{align}
h^0(\cO_\cP(1,3,1))=48\,,\qquad h^0(\cN_{2}^*\otimes\cO_\cP(1,3,1))=h^0(\cO_\cP(1,1,0))=4 \ .
\end{align}
Hence, upon imposing the second of the four complete intersection equations, 4 of the 48 monomials become trivial. Similarly, for the third and fourth entry in the monad map, we observe that
\begin{align}
\begin{split}
h^0(\cO_\cP(2,3,1))=72\,,\qquad h^0(,\cN_{1}^*\otimes\cO_\cP(2,3,1))=h^0(\cO_\cP(0,3,0))=4\,,\\ 
h^0(\cN_{2}^*\otimes\cO_\cP(2,3,1))=h^0(\cO_\cP(2,1,0))=6\,.
\end{split}
\end{align}
Therefore, 4 ambient space monomials become trivial on the first hypersurface and 6 on the second, for a total of 10 on the CICY (which is the intersection of all four ambient space hypersurfaces).

To identify further over-parametrizations, we next look at the bundle endomorphisms $A^*\otimes A$ and $B^* \otimes B$. Since $r_B=1$, we have $B^*\otimes B={\cal O}_X$, which corresponds to removing one overall scaling degree of freedom. For the $r_A\times r_A$ cohomology matrix $h^0(X,{\cal O}_X(a_\alpha-a_{\alpha'}))$ we find
\begin{align}
h^0(X,{\cal O}_X(a_\alpha-a_{\alpha'}))=
\left(
\begin{array}{llll}
 1 & 8 & 0 & 0 \\
 0 & 1 & 0 & 0 \\
2 & 14 & 1 & 1 \\
2 & 14 & 1 & 1 \\
\end{array}
\right)
\end{align}
As can be seen, this bundle has several automorphisms. The diagonal entries are unity and, hence, correspond to overall scaling symmetries.  The bottom right $2\times2$ block arises from the fact that $a_3=a_4$.

We find that all other cohomology dimension in Eq.~\eqref{eq:MonadBundleEndDim} are zero. In particular, there are no further contributions to $h^{1}(X,V\otimes V^*)$ that do not descend from the ambient space, which means that the bundle is bundle favorable and does not under-parametrize the bundle moduli space. Putting everything together, we find that we need to subtract 46 parameters that over-parametrize the bundle moduli space due to symmetries in the line bundle sums $A$ and $B$. This leaves precisely $176-46=130$ bundle moduli on $X$ which is the correct number.

\subsubsection*{The Pfaffians}

In order to compute the Pfaffians, we next tensor the monad with $\mathcal{O}_X(-1,0,0)$ and restrict to the first $\mathbbm{P}^1$ to obtain
\begin{align}
0\xrightarrow{~}V_i\xrightarrow{~}\mathcal{O}_{\mathbbm{P}^1}\oplus\mathcal{O}_{\mathbbm{P}^1}\oplus\mathcal{O}_{\mathbbm{P}^1}(-1)\oplus\mathcal{O}_{\mathbbm{P}^1}(-1)\xrightarrow{f|_{\gamma_i}}\mathcal{O}_{\mathbbm{P}^1}(1)\xrightarrow{~}0\,.
\end{align}
Next, we look at the sections $H^{0}(V_i)$. Since the line bundle $\mathcal{O}_{\mathbbm{P}^1}(-1)$ does not have any sections, we drop it from the subsequent discussion and compute
\begin{align}
H^{0}(V_i) \cong\text{Ker}\left[f|_{\gamma_i}:~H^{0}\left(\mathcal{O}_{\mathbbm{P}^1}^{\oplus2})\rightarrow H^0(\mathcal{O}_{\mathbbm{P}^1}(1))\right)\right]\,,
\end{align}
where $i=1,\ldots,8$ labels the eight curves in the curve class $\gamma$ of the first $\mathbbm{P}^1$. For the restricted monad map we find, after dropping the last two columns that correspond to the $\mathcal{O}_{\mathbbm{P}^1}(-1)$ terms and inserting the position $y_{i}$ of the $i^\text{th}$ curve, that
\begin{align}
\label{eq:restricted}
f(\vec{z}_1,\vec{z}_2,\vec{z}_3)|_{C_i}=\left(
\begin{array}{c}
f_{1,3,1}(z_{1,0},z_{1,1};y_{i}) \\ f_{1,3,0}(z_{1,0},z_{1,1};y_{i})
\end{array}
\right)
=
\left(
\begin{array}{c}
z_{1,0} \;p_{3,1}(y_i)+z_{1,1} \;q_{3,1}(y_i) \\ z_{1,0} \;s_{3,0}(y_i) + z_{1,1} \;t_{3,0}(y_i)
\end{array}
\right)\,,
\end{align}
with 
\begin{align}
p_{3,1}(y_{i})=\sum_{r=1}^{24}\beta_r^{(1)} m_{3,1}^r(y_{i})\,,\qquad q_{3,1}(y_{i})=\sum_{r=1}^{24}\beta_r^{(2)} m^r_{3,1}(y_{i})\,,\\
s_{3,0}(y_{i})=\sum_{r=1}^{4}\beta_r^{(3)} m_{3,0}^r(y_{i})\,,\qquad t_{3,0}(y_{i})=\sum_{r=1}^{4}\beta_r^{(4)} m^r_{3,0}(y_{i})\,,
\end{align}
and $m_{d_1,d_2}^r(y)=m_{d_1,d_2}^r(\vec{z}_3,\vec{z_3})$ denoting monomials of multidegree $(d_2,d_3)$ in $(\vec{z}_2,\vec{z}_3)$. The coefficients $\beta_r^{(i)}$ parametrize the bundle moduli space of the (ambient space) monad bundle. We could remove the redundancies as explained above, but the result does not depend on this.

Choosing $(\alpha,\beta)^T$ with $\alpha,\beta\in\mathbbm{C}$ as a basis of $\mathcal{O}_{\mathbbm{P}^1}^{\oplus2}$, this means
\begin{align}
\label{eq:kerCondition}
\left(\begin{array}{c}\alpha\\\beta\end{array}\right) \in \text{ker}\left[f|_{\gamma_i}\right] ~~~\text{if}~~~ (x_{1,0}~x_{1,1})\left(\begin{array}{ccc}
p_{3,1}(y_i) & q_{3,1}(y_i)\\
s_{3,0}(y_i) & t_{3,0}(y_i)
\end{array}\right)\left(\begin{array}{c}\alpha\\\beta\end{array}\right)=0\,.
\end{align}

The kernel of $f|_{\gamma_i}$ will be non-trivial if the determinant $\delta$ of the map in \eqref{eq:kerCondition} vanishes. So to test the Beasley-Witten vanishing result, we compute
\begin{align}
\delta_i=p_{3,1}(y_i) t_{3,0}(y_i) - q_{3,1}(y_i) s_{3,0}(y_i)\,.
\end{align} 
By construction, the $\delta_i$'s we obtain this way are polynomials in the vector bundle moduli $\beta_r^{(i)}$. As we have mentioned above, the functional form of the $\delta_i$ is very involved and will not be presented here.
The next step is to check whether
\begin{align}
\label{eq:CICY7735BW}
\sum_{i=1}^8 \lambda_i \delta_i\stackrel?=0 
\end{align}
for some $\lambda_i\in\mathbbm{C}^*$. We find that there are 192 different monomials bilinear in the bundle moduli, and the $\delta_i$ form linearly independent combinations of these, that is, no $\lambda_i$ exist to make the above sum vanish. We conclude that the instanton contributions cannot cancel and, hence, that the non-perturbative superpotential $W_\gamma$ is non-vanishing.


\subsection{An examples with an SU(3) extension bundle}
\label{sec:Extension_Example}


In this section, we will present an example based on an extension bundle with $SU(3)$ structure group.

\subsubsection*{The geometry}
We choose CICY number 7860 in the list of Ref.~\cite{Candelas:1987kf}. Its configuration matrix reads
\begin{align}
\label{eq:ExtensionExampleCICYConf}
X\sim\left.\left[
\begin{array}{c|ccc}
\mathbbm{P}^{1}&1&1&0\\
\mathbbm{P}^{1}&0&0&2\\
\mathbbm{P}^{2}&1&0&2\\
\mathbbm{P}^{2}&0&1&2\\
\end{array}
\right]^{(4,68)}\quad \right| \quad
\begin{array}{l}
\vec{z}_1=[z_{1,0}:z_{1,1}]\\
\vec{z}_2=[z_{2,0}:z_{2,1}]\\
\vec{z}_3=[z_{3,0}:z_{3,1}:z_{3,2}]\\
\vec{z}_4=[z_{4,0}:z_{4,1}:z_{4,2}]\,.
\end{array}
\end{align}

We define the bundle $V$ as the extension $0\rightarrow A\rightarrow V\rightarrow B\rightarrow 0$ of line bundle sums
\begin{align}
\label{eq:V_vis}
A=\cO_X(-2, 3, -1, 1)\,,\qquad B=\cO_X(0, 0, 2, -2)\oplus \cO_X(2, -3, -1, 1)\,.
\end{align}

\subsubsection*{Consistency conditions}
We can check that the bundle satisfies the usual consistency constraints, namely that $c_1 (V)=0$ and that $c_2 (TX) - c_2 (V)$ is effective. Furthermore,  we have checked that the trivial extension $A\oplus B$ allows for a common slope zero locus for all line bundles.

In order to check that $V $ is a non-trivial extension, we compute the dimensions of the cohomologies following the techniques described in~\cite{CICYPackage}. We find that
\begin{align}
\label{eq:CohomDimensions}
\begin{split}
h^\bullet({\cal O}_\cP(a_1-b_1))&=(0, 0, 0, 40, 0, 0, 0)\,,\qquad h^\bullet({\cal O}_X(a_1-b_1))=(0, 16, 0, 0)\,,\\
h^\bullet({\cal O}_\cP(a_1-b_2))&=(0, 21, 0, 0, 0, 0, 0)\,,\qquad h^\bullet({\cal O}_X(a_1-b_2))=(0, 21, 25, 0)\,.
\end{split}
\end{align}
Since $h^1({\cal O}_X(a_1-b_1))$ and $h^1({\cal O}_X(a_1-b_2))$ are non-zero, the bundle has non-trivial extensions.

\subsubsection*{Finding the curves}
The CICY geometry for the first $\mathbbm{P}^1$ is of type I, and the location of the two curves in transverse space $\mathcal{Q}$ are given by the complete intersection
\begin{align}
\label{eq:ExtensionExampleCurves}
\{y_i\}=\left[
\begin{array}{c|ccccc}
\mathbbm{P}^{1}&0&0&0&0&2\\
\mathbbm{P}^{2}&1&1&0&0&2\\
\mathbbm{P}^{2}&0&0&1&1&2\\
\end{array}
\right]\,,
\end{align}
with $y=(\vec{z}_2,\vec{z}_3,\vec{z}_4)$. It is easy to see this configuration describes two points: the first two equations fix a point in the first $\bP^2$, the next two equations fix a point in the second $\bP^2$, and the last equation, being quadratic in $\vec{z}_2$, leads to two points in the $\bP^1$. We denote these points by $y_i$, and the corresponding curves by $\gamma_i$, where $i=1,2$. Note that, since the coefficients in the defining polynomials~\eqref{eq:ExtensionExampleCICYConf} parametrize (in a redundant way) the complex structure moduli space, these points depend on the complex structure. We have also cross-checked our results by computing the Gromov-Witten invariants, following the methods of Ref.~\cite{Hosono:1993qy}.

\subsubsection*{Parametrizing the bundle moduli space}
To  compute the Pfaffians  we first need to explicitly parametrize the bundle moduli space.  As calculated in~\eqref{eq:CohomDimensions}, the dimensions of the corresponding bundle extensions on $X$ are 16 and 21, respectively. As it turns out, the Pfaffians only depend on the moduli of the latter. 
This is helpful, since the former do not descend from $H^1$ of the ambient space, since $h^1({\cal O}_\cP(a_1-b_1))=0$ in~\eqref{eq:CohomDimensions}. 
Let us denote the projection from $\cP$ onto $\mathcal{Q}$ by~$\pi_{\mathcal{Q}}$,
\begin{align}
\begin{split}
\pi_{\mathcal{Q}}:\cP&~\rightarrow~\mathcal{Q}\,,\\
\bP^1\times\bP^1\times\bP^2\times\bP^2&~\mapsto~\bP^1\times\bP^2\times\bP^2\,.
\end{split}
\end{align}

The relevant extension space is then
\begin{align}
H^1({\cal O}_\cP(a_1-b_2))=H^1(\cO_\cP(4,6,0,0))=H^1(\cO_{C_i}(-4))\otimes H^1(\cO_{\mathcal{Q}}(6,0,0))\,,
\end{align}
where we have used the K\"unneth and Bott formulas. Due to Serre duality, we have
\begin{align}
\label{eq:H1Om4}
H^1(O_{C_i}(-4))\simeq H^0(O_{C_i}(2))^*\,,
\end{align}
and this space is three-dimensional. Since we will need it later, we first introduce a basis 
$\{t_0,t_1\}$ for $H^0(O_{\gamma_i}(1))^*$ and the dual basis $\{r_0,r_1\}$ for $H^1(O_{\gamma_i}(-3))$. A natural basis for the left-hand side of~\eqref{eq:H1Om4} is then $\{r_0^2,r_0 r_1,r_1^2\}$, dual to the degree 2 polynomials of the right-hand side $\{t_0^2,t_0 t_1,t_1^2\}$. An arbitrary element $v\in H^1(\cO_\cP(-4,6,0,0))$ can then be written as
\begin{align}
\label{eq:H1Element}
v=r_0^2f^{(1)}_{6,0,0}(\vec{z}_2,\vec{z}_3,\vec{z}_4)+r_0 r_1 f^{(2)}_{6,0,0}(\vec{z}_2,\vec{z}_3,\vec{z}_4)+ r_1^2f^{(3)}_{6,0,0}(\vec{z}_2,\vec{z}_3,\vec{z}_4) \ .
\end{align}
That is, the three polynomials $f^{(i)}$ are homogeneous polynomials of degree 6 in $[z_{2,0}:z_{2,1}]$,
\begin{align}
\begin{split}
f^{(1)}_{6,0,0}(\vec{z}_2,\vec{z}_3,\vec{z}_4) &= \sum_{i=0}^6 \beta_i^{(1)} z_{2,0}^i z_{2,1}^{6-i}\,,\qquad
f^{(2)}_{6,0,0}(\vec{z}_2,\vec{z}_3,\vec{z}_4) = \sum_{i=0}^6 \beta_i^{(2)} z_{2,0}^i z_{2,1}^{6-i}\,,\\
f^{(3)}_{6,0,0}(\vec{z}_2,\vec{z}_3,\vec{z}_4) &= \sum_{i=0}^6 \beta^{(3)}_i z_{2,0}^i z_{2,1}^{6-i}\,.
\end{split}
\end{align}
The $3\times7=21$ coefficients $\beta_i^{(r)}$ precisely parametrize the 21-dimensional extension space.

\subsubsection*{Computing the Pfaffians}
Now that we have a parametrization of bundle moduli space we can move on to computing the Pfaffians. We start from the extension~\eqref{eq:V_vis} and tensor it with $\cO_\cP(-1,0,0,0)$ to obtain
\begin{align}
0\rightarrow\cO_\cP(-3,3,-1,1)\rightarrow \mathcal{V}\otimes\cO_\cP(-1,0,0,0)\rightarrow\cO_\cP(-1,0,2,-2)\oplus\cO_\cP(1,-3,-1,1)\,.
\end{align}
Now we take the direct image with the projection $\pi_\mathcal{Q}$, which leads to the long exact sequence
\small
\begin{align}
\begin{array}{l@{\;}l@{\;}l@{\;}l@{\;}}
0&\rightarrow\phantom{R^1}\pi_{\mathcal{Q}^*}\cO(-3,3,-1,1)&\rightarrow\phantom{R^1}\pi_{\mathcal{Q}^*}(\mathcal{V}\otimes\cO_\cP(-1,0,0,0))&\rightarrow
\phantom{R^1}\pi_{\mathcal{Q}^*}(\cO_\cP(-1,0,2,-2)\oplus\cO_\cP(1,-3,-1,1))\\
  &\rightarrow R^1\pi_{\mathcal{Q}^*}\cO(-3,3,-1,1)&\rightarrow R^1\pi_{\mathcal{Q}^*}(\mathcal{V}\otimes\cO_\cP(-1,0,0,0))&\rightarrow R^1\pi_{\mathcal{Q}^*}(\cO_\cP(-1,0,2,-2)\oplus\cO_\cP(1,-3,-1,1))\\
  & \rightarrow0&&\,.
\end{array}
\end{align}
\normalsize
At each point in $\mathcal{Q}$ and for any line bundle $L$, $\pi_\mathcal{Q} L$ and $R^1\pi_\mathcal{Q} L$ are generated by the zeroth and first cohomology group of the fiber (i.e.,\ the first ambient space $\bP^1$ factor). Since
\begin{align}
\begin{split}
H^0(\cO_{\bP^1}(-3))&=0\,,\qquad H^0(\cO_{\bP^1}(-1))=0\,,\\ 
H^1(\cO_{\bP^1}(-1))&=0\,,\qquad H^1(\cO_{\bP^1}(1))=0\,,
\end{split}
\end{align}
and, by Bott's formula,
\begin{align}
R^1\pi_{\mathcal{Q}^*}\cO(-3,3,-1,1)=H^1(\cO_{\bP^1}(-3))\otimes\cO_\mathcal{Q}(3,-1,1)\,,
\end{align}
the long exact sequence becomes
\begin{align}
\label{eq:LESPfaffian}
\begin{array}{l@{\;}l@{\;}l@{\;}}
0&\xrightarrow{\phantom{\delta V}}\phantom{R^1}\pi_{\mathcal{Q}^*}(\mathcal{V}\otimes\cO_\cP(-1,0,0,0))&\xrightarrow{\phantom{\delta V}} H^0(\cO_{\bP^1}(1))\otimes\cO_\mathcal{B}(-3,-1,1))\\
  &\xrightarrow{\;f~} H^1(\cO_{\bP^1}(-3))\otimes\cO_\mathcal{Q}(3,-1,1))&\xrightarrow{\phantom{\delta V}} R^1\pi_{\mathcal{Q}^*}(\mathcal{V}\otimes\cO_\cP(-1,0,0,0))\\
  &\xrightarrow{\phantom{\delta V}} 0\,.&
\end{array}
\end{align}
Since $h^0(\cO_{\bP^1}(1))=2=h^1(\cO_{\bP^1}(-3))$, the map $f$ can be represented by a $2\times2$ matrix. We can now restrict the exact sequence~\eqref{eq:LESPfaffian} to the curves $\gamma_i$, $i=1,2$, by considering it at the two points $(\vec{z}_{2*}^{\;i},\vec{z}_{3*}^{\;i},\vec{z}_{4*}^{\;i})\subset \mathcal{Q}$, $i=1,2$. We see that the first term in the sequence becomes
\begin{align}
\pi_{\mathcal{Q}^*}(\mathcal{V}\otimes\cO_\cP(-1,0,0,0))=H^0(V_{i})\,.
\end{align}
As discussed in Section~\ref{sec:Introduction}, this is precisely the space of zero modes of the Dirac operator. Consequently, this will be non-trivial if $f$ has a non-trivial kernel,
\begin{align}
\text{Pfaff}(\overline{\partial}_{V_i})=0\quad\Leftrightarrow\quad \text{det}(f)=0\,.
\end{align}
The map $f$ is simply given by multiplication by $v\in H^1(\cO_\cP(-4,6,0,0))$; see for example~\eqref{eq:H1Element}. It is constructed by acting on $v$ with the basis elements $\{t_0,t_1\}$ introduced above~\eqref{eq:H1Element},
\begin{align}
\begin{split}
v(t_0)&=r_0 f_{6,0,0}^{(1)}(\vec{z}_2,\vec{z}_3,\vec{z}_4)+r_1 f_{6,0,0}^{(1)}(\vec{z}_2,\vec{z}_3,\vec{z}_4)\,,\\[4pt]
v(t_1)&=r_0 f_{6,0,0}^{(2)}(\vec{z}_2,\vec{z}_3,\vec{z}_4)+r_1 f_{6,0,0}^{(3)}(\vec{z}_2,\vec{z}_3,\vec{z}_4)\,.
\end{split}
\end{align}
Hence,
\begin{align}
\begin{split}
f&=\begin{pmatrix}
 f_{6,0,0}^{(1)}(\vec{z}_2,\vec{z}_3,\vec{z}_4) &  f_{6,0,0}^{(2)}(\vec{z}_2,\vec{z}_3,\vec{z}_4)\\
 f_{6,0,0}^{(2)}(\vec{z}_2,\vec{z}_3,\vec{z}_4) &  f_{6,0,0}^{(3)}(\vec{z}_2,\vec{z}_3,\vec{z}_4)
\end{pmatrix}
\,,\\[8pt]
\delta&=\text{det}(f)=f_{6,0,0}^{(1)}(\vec{z}_2,\vec{z}_3,\vec{z}_4) f_{6,0,0}^{(3)}(\vec{z}_2,\vec{z}_3,\vec{z}_4)-[f_{6,0,0}^{(2)}(\vec{z}_2,\vec{z}_3,\vec{z}_4)]^2\,.
\end{split}
\end{align}
Thus, the Pfaffian for the curve $\gamma_i$ is proportional to $\delta_i$, that is, $\delta$ evaluated at $y_i$.

\subsubsection*{Checking Beasley-Witten cancellation}
We find that a necessary condition for the vanishing of the superpotential contribution is thus
\begin{align}
\label{eq:PfaffianVanishing}
\lambda_1\left(f_{6,0,0}^{(1)}(y_1) f_{6,0,0}^{(3)}(y_1)-[f_{6,0,0}^{(2)}(y_1)]^2\right) + \lambda_2 \left(f_{6,0,0}^{(1)}(y_2) f_{6,0,0}^{(3)}(y_2)-[f_{6,0,0}^{(2)}(y_2)]^2\right)=0
\end{align}
for some $\lambda_1,\lambda_2\in\mathbbm{C}$. Note that~\eqref{eq:PfaffianVanishing} is quadratic in the 21 bundle moduli or, more precisely, it contains terms of 
the form $\alpha_i\gamma_j$ and $\beta_i\beta_j$. We find that for no values of $\lambda_1,\lambda_2$ can eq.~\eqref{eq:PfaffianVanishing} hold ,which means that the instanton contribution 
\begin{align}
\label{eq:PfaffianNonVanishing}
\lambda_1\left(f_{6,0,0}^{(1)}(y_1) f_{6,0,0}^{(3)}(y_1)-[f_{6,0,0}^{(2)}(y_1)]^2\right) + \lambda_2 \left(f_{6,0,0}^{(1)}(y_2) f_{6,0,0}^{(3)}(y_2)-[f_{6,0,0}^{(2)}(y_2)]^2\right)
\end{align}
is non-vanishing. 


\subsection{An example admitting both extension and monad descriptions}
\label{sec:Extension_Monad}


In this subsection, we consider an example involving an extension bundle for which we can find an equivalent monad description,
following our discussion in subsection~\ref{sec:ExensionMonadMap}. This means we can compute the Pfaffians in two different ways. 
We find both cases lead to the same results; thus providing a non-trivial consistency check of our technique. 


\subsubsection*{The geometry}

We study the CICY with configuration matrix
\begin{align}
\label{eq:CICYConf}
X\in\left.\left[
\begin{array}{c|cccc}
\bP^1 & 1 & 1 & 0 & 0\\
\bP^1 & 1 & 0 & 1 & 0\\
\bP^1 & 1 & 0 & 0 & 1\\
\bP^2 & 0 & 1 & 1 & 1\\
\bP^1 & 2 & 0 & 0 & 0\\
\bP^1 & 2 & 0 & 0 & 0\\
\end{array}
\right]^{(6,54)}\quad \right| \quad
\begin{array}{l}
\vec{z}_1=[z_{1,0}:z_{1,1}]\\
\vec{z}_2=[z_{2,0}:z_{2,1}]\\
\vec{z}_3=[z_{3,0}:z_{3,1}]\\
\vec{z}_4=[z_{4,0}:z_{4,1}:z_{4,2}]\\
\vec{z}_5=[z_{5,0}:z_{5,1}]\\
\vec{z}_6=[z_{6,0}:z_{6,1}]~\ .
\end{array}
\end{align}
This manifold is an ineffective split, leading to a favorable configuration, of CICY 7709 in the list of Ref.~\cite{Candelas:1987kf}. The first four ambient space factors $(\bP^1)^3 \times \bP^2$ together with the first three equations, define a (non-generic) dP$_3$ where all three blow-ups are co-linear \cite{Green:1987zp}. 
The second Chern classes are
\begin{align}
\label{eq:c2TX}
c_{2,i}=(24,24,24,36,24,24)\,,
\end{align}
and the Gromov-Witten invariants in the class of the $i^{\text{th}}$ ambient space factor are
\begin{align}
n_{1,0,0,0,0,0}=8\,,\;
n_{0,1,0,0,0,0}=8\,,\;
n_{0,0,1,0,0,0}=8\,,\;
n_{0,0,0,1,0,0}=0\,,\;
n_{0,0,0,0,1,0}=36\,,\;
n_{0,0,0,0,0,1}=36\,.
\end{align}
We have computed these with our method~\cite{Buchbinder:2017azb} (where applicable) and have cross-checked and supplemented with the method of Ref.~\cite{Hosono:1993qy}.

Similar to Refs.~\cite{Buchbinder:2016rmw,Buchbinder:2017azb}, we consider an \SU3 double extension bundle $V$ defined via three line bundle sums $A$, $B$, $C$, as
\begin{align}
\label{eq:ExensionSequence}
\begin{split}
0\rightarrow A \rightarrow V' \rightarrow B \rightarrow 0\,,\\
0\rightarrow V'\, \rightarrow V \rightarrow C\rightarrow 0\,.
\end{split}
\end{align}
The bundles $A$, $B$, $C$ consist of a single line bundle each and are given by
\begin{align}
A=\cO_X(-2,2,0,0,1,1)\,,\quad B=\cO_X(0,1,0,0,-1,0)\,,\quad C=\cO_X(2,-3,0,0,0,-1)\,.
\label{e6}
\end{align}
We will focus on the instanton contributions of the first ambient space factor.


\subsubsection*{Bundle cohomologies}
We start by listing several bundle cohomologies that will be useful here and later.
\begin{subequations}
\begin{align}
h^\bullet(\cA\otimes \cB^*)	&=(0,12,0,0,0,0,0,0)\,,		\quad  &h^\bullet(A\otimes B^*)	&=(0,12,0,0)\,,	\label{eq:coh12}\\
h^\bullet(\cA\otimes \cC^*)	&=(0,108,0,0,0,0,0,0)\,,		\quad  &h^\bullet(A\otimes C^*)	&=(0,108,0,0)\,,	\label{eq:coh13}\\
h^\bullet\cB\otimes \cC^*)	&=(0,0,0,0,0,0,0,0)\,,			\quad  &h^\bullet(B\otimes C^*)	&=(0,0,0,0)\,.		\label{eq:coh23}
\end{align}
\end{subequations}

\subsubsection*{Non-trivial extension}
\label{sec:NontrivialExtension}
First we check that the extension is non-trivial, i.e.,\ we need to check ${\rm Ext}(B,A)\cong H^1(X,A\otimes B^*)$ and ${\rm Ext}(C,V)\cong H^1(X,V'\otimes C^*)$. The former is given directly by Eq.~\eqref{eq:coh12} and its dimension is $12$. To compute the latter, we tensor the first line of~\eqref{eq:ExensionSequence} by $C$ and look at the resulting long exact sequence in cohomology,
\begin{align}
\begin{split}
0\rightarrow 	& H^0(A \otimes C^*) 			\rightarrow H^0(V' \otimes C^*) 							\rightarrow H^0(B \otimes C^*) 	\rightarrow\\
						& H^1(A \otimes C^*) 			\rightarrow \underline{H^1(V' \otimes C^*)} 	\rightarrow H^1(B \otimes C^*) 	\rightarrow\\
						& H^2(A \otimes C^*) 			\rightarrow \ldots
\end{split}
\end{align}
We are interested in the underlined term. Since $h^0(B \otimes C^*)=h^1(B \otimes C^*)=0$, we find that 
\begin{align}
h^1(A \otimes C^*) = h^1(V' \otimes C^*) = 108\,.
\end{align}
Hence, there exist non-trivial extensions.

\subsubsection*{Bianchi identities}
The Bianchi identities can be satisfied provided $c_2(TX) - c_2(V)$ is effective. Comparing the second Chern class
\begin{align}
\label{eq:ch2VExt}
c_2(V) = (14,- 10, 2, 3,-18, -22)\,,
\end{align}
with that of the tangent bundle in Eq.~\eqref{eq:c2TX} shows that the difference is indeed effective.

\subsubsection*{Bundle stability}
\label{sec:Bundle_stability}
Verifying poly-stability of an \SU3 bundle on a manifold with $h^{1,1}=6$ is very involved. So we will content ourselves with the following checks:
\begin{itemize}
\item Note that $A$ injects into $V'$ and $V'$ injects into $V$. We check that neither injecting subsheaf destabilizes $V$.
\item There exist a common slope zero locus for the trivial extension $V=A\oplus B \oplus C$.
\item Turing on vevs for bundle moduli that lead away from the trivial extension still preserve the D-term equations.
\end{itemize}
We verify these requirements numerically with Mathematica. 

Let us be more explicit about the last check based on the D-terms. The process of constructing a non-trivial
extension can be described in supersymmetric low-energy field theory and is equivalent to finding supersymmetric vacua
of the D-term equations~\cite{Anderson:2009sw, Anderson:2009nt}. 
We expand the K\"ahler form
\begin{align}
J=\sum_{i=1}^6 t_i D_i\,,
\end{align}
where $t_i$ are the K\"ahler parameters and $D_i$ are the (duals of the) divisors that are obtained from pulling back the hyperplane class of the $i^\text{th}$ ambient space factor $\bP^{n_i}$ to $X$. For notational convenience we also define 
\begin{align}
v_i = \sum_{j,k} d_{ijk} t_j t_k\,,
\label{e3}
\end{align}
where $d_{ijk}$ are the triple intersection numbers. The $v_i$ are proportional to the volume of divisor $D_i$ and thus need to be positive. At the split locus, we get three D-term equations (one for each line bundle), out of which two are linearly independent (due to the fact that the bundle is an S[U(1)$^3$] bundle). The bundle moduli at the split locus are singlets under the non-Abelian groups but they do carry U(1) charge. Their multiplicity is counted by $h^\bullet(L_i\otimes L_j^*)$ (where $L_i\in\{A,B,C\}$) and they carry charge $1$ and $-1$ under the $i^\text{th}$ and $j^\text{th}$ U(1), respectively. We denote the matter fields by $M_{(q_1,q_2,q_3)}$. From the cohomologies \eqref{eq:coh12}-\eqref{eq:coh23} we see that there are 12 fields $M_{(1,-1,0)}$ and 108 fields $M_{(1,0,-1)}$. The D-term equations can then be written as
\begin{align}
\begin{split}
2 v_1+v_3-2 v_4+v_5-\sum_{i=1}^{12}|\langle M_{(1,-1,0)}^i\rangle|^2 -\sum_{i=1}^{108}|\langle M_{(1,0,-1)}^i \rangle|^2&=0\,,\\
v_1-v_3+\sum_{i=1}^{12}|\langle M_{(1,-1,0)}^i\rangle|^2&=0\,,\\
-3 v_1+2 v_4-v_5+\sum_{i=1}^{108}|\langle M_{(1,0,-1)}^i\rangle|^2&=0\,.
\end{split}
\end{align}
We observe that these equations do have 
solutions
for positive K\"abler parameters  $t_i$ (which depend on $v_i$ as in eq.~\eqref{e3})
and non-negative vevs $|M_{(q_1,q_2,q_3)}|^2$. 
The fact that they have solutions where all vevs are turned off means that the bundle is stable at the split locus. Furthermore, 
there are many flat directions in the combined K\"ahler and bundle moduli space, which correspond to supersymmetric deformations away from the trivial extension.


\subsubsection*{Checking Beasley-Witten cancellation}
\label{sec:BWCancellation}
In this section, we check Beasley-Witten cancellation. The computation is analogous to the one presented in 
subsection~\ref{sec:Extension_Example}, so we will be brief here. 
As we have shown above the (relevant part of the) bundle moduli space of $V$ is given by $h^1(\cA \otimes \cC^*)=108$, where
\begin{align}
\cA \otimes \cC^* = \mathcal{O}_\cP(-4,5,0,0,1,2)\,.
\end{align}
This allows us to write the extension in terms of ambient space coordinates. Using Serre duality, we can write an arbitrary element $m\in H^1(\cA \otimes \cC^*)$ as 
\begin{align}
\label{eq:ExtBundleModuliMap}
m=z_{1,0}^2 f^{(1)}_{(5,1,2)}(\vec{z}_2,\vec{z}_5,\vec{z}_6) + z_{1,0} z_{1,1} f^{(2)}_{(5,1,2)}(\vec{z}_2,\vec{z}_5,\vec{z}_6) + z_{1,1}^2 f^{(3)}_{(5,1,2)}(\vec{z}_2,\vec{z}_5,\vec{z}_6)\,.
\end{align}
The subscript indicates that the three functions $f^{(i)}$ are homogeneous of degree $(5,1,2)$ in the $(\bP^1)^3$ coordinates $(\vec{z}_2,\vec{z}_5,\vec{z}_6)$. There are 36 monomials of this multi-degree, so we have a total of 108 parameters in $m$. These 108 parameters parametrize the $h^1(\cA\otimes \cC^*)=108$ dimensional extension space.  In fact, the moduli space is only 107-dimensional since we need to projectivize,

Next, we find the 8 curves in the curve class of the first ambient space $\bP^1$ factor following the steps outlined for a type I geometry in Section~\ref{sec:CY}. The 8 curves will fix 8 distinct points in the transverse space $\mathcal{Q}$. We find that the Pfaffian is proportional to the map $\delta$ with
\begin{align}
\label{eq:PfaffianExtensionBundle}
\delta = f^{(1)}_{(5,1,2)}(\vec{z}_2,\vec{z}_5,\vec{z}_6) f^{(3)}_{(5,1,2)}(\vec{z}_2,\vec{z}_5,\vec{z}_6) -[f^{(2)}_{(5,1,2)}(\vec{z}_2,\vec{z}_5,\vec{z}_6)]^2\,.
\end{align}
By substituting the values of $y_i=(\vec{z}_2,\vec{z}_3,\vec{z}_4,\vec{z}_5,\vec{z}_6)_i$, $i=1,\ldots,8$, that give the position of the 8 curves, we obtain 
8 polynomials in the 108 bundle moduli parameters for arbitrary complex structure. 
Each polynomial has 1962 terms and we find that the 8 polynomials are linearly independent. 
Hence, there cannot be any cancellation among them and, consequently, the instanton superpotential is non-zero.


\subsubsection*{An equivalent monad bundle description}
\label{sec:Monad7709}
Next, we apply the method explained in Section~\ref{sec:ExensionMonadMap} to find a monad bundle description for the previous extension bundle. 
We need to choose the line bundle sums $\tilde{A}_a$ and $\tilde{B}_a$ in Eq.~\eqref{extmon} such that the bundle map $F$ allows for non-trivial 
deformations so that we get a bona fide \SU3 monad bundle rather than a (partially) split bundle. Also, the construction requires $\mathbbm{P}^1$ factors, which is 
why we chose the extension bundle to be the trivial bundle on the $\bP^2$ factor. In this way, we obtain an infinite family of monad bundles
\begin{align}
\begin{split}
0\rightarrow V &\rightarrow\cO_X(r-2,2,0,0,1,1)\oplus\cO_X(\tilde{r}-2,2,0,0,1,1)\oplus\cO_X(0,1,0,0,-1,0)\oplus\cO_X(2,-3,0,0,0,-1)\\
&\xrightarrow{f} \cO_X(r+\tilde{r}-2,2,0,0,1,1)\rightarrow 0\,,
\label{e4}
\end{split}
\end{align}
with $r,\tilde{r}\geq2$. The monad map $f$ can be written as a $4\times1$ matrix
\begin{align}
f=\left(f^{(1)}_{(\tilde{r}, 0, 0, 0, 0, 0)},\;f^{(2)}_{(r, 0, 0, 0, 0, 0)},\;f^{(3)}_{(r+\tilde{r}-2, 1, 0, 0, 2, 1)},\; f^{(4)}_{(r+\tilde{r}-4, 5, 0, 0, 1, 2)}\right)\,,
\end{align}
where the subscripts indicate the multi-degrees of the polynomials. Note that the monad and the extension bundle are only equivalent on the CY. On the ambient space, their moduli space has, in general, different dimensions. (This is possible since some of the contributions to the bundle moduli come from higher terms in the Koszul sequence.)

As a cross-check, we have verified that the cohomologies $h^{\bullet}(V)$ for the extension bundle and its associated monad agree. Moreover, we have computed the number of singlets from $h^{\bullet}(V^* \otimes V)=(1,118,118,1)$. The 118 bundle moduli match precisely the moduli of the extension bundle. 
The extension of $V'$ has $h^1(A\otimes B^*)=12$, and due to projectivization, these correspond to 11 moduli. Similarly, the extension of $V$ has $h^1(V'\otimes C^*)=108$, 
which means that the extension space is 107-dimensional. Together, we find $11+107=118$ moduli. These correspond to the singlets $M_{(1,-1,0)}^i$ and $M_{(1,0,-1)}^i$, respectively. 

Let us next illustrate how different choices of $r$ and $\tilde{r}$ can be used to make the monad bundle favorable so that all bundle moduli descend from the ambient space. When counting the number of parameters in the monad map $f$, we find
\begin{align}
\label{eq:NumberOfParametersMonad}
\text{number of parameters}=(r+1)+(\tilde{r}+1)+12(r+\tilde{r}-1)+36(r+\tilde{r}-3)=49(r+\tilde{r})-118~\ .
\end{align}
For the minimum value $r=\tilde{r}=2$, the monad map has only 78 parameters, and some of them are not even independent. So, not all of the 118 bundle moduli are explicit realized by the monad in this case. For $r+\tilde{r}\geq5$, there are more than 118 bundle moduli in the monad map, so we have over-parametrization. By looking at the cohomologies, the minimal choice which leads to $\geq 118$ parameters in the monad map is $r=\tilde{r}=3$.

In more detail, we look at the various contributions from~\eqref{eq:MonadBundleEndDim}. The symmetries of the bundle we need to subtract come from $h^0(B\otimes B^*)$ and $h^0(C\otimes C^*)$. In the present case, we can write them as $4\times4$ and $1\times 1$ matrices, respectively. These matrices will have a 1 on the diagonal corresponding to scaling symmetries of the line bundle. The off-diagonal entries $(\beta,\beta')$ correspond to maps from the line bundle ${\cal O}_X(b_\beta')$ to the line bundle in ${\cal O}_X(b_\beta)$. The dimensions are
\begin{align}
h^0({\cal O}_X(b_\beta -b_{\beta'}))=\begin{pmatrix}
1&1&12(r-1)&0\\
1&1&12(\tilde{r}-1)&0\\
0&0&1&0\\
0&0&0&1\\
\end{pmatrix}
\,,\quad
h^0(C\otimes C^*)=1
\,.
\end{align}
Hence, the overparametrization due to symmetries in $B$ and $C$ is $12(r+\tilde{r})-18$. Just removing these redundancies (there are also redundancies coming from $h^1(B\otimes C^*)$) from~\eqref{eq:NumberOfParametersMonad}, the number of independent bundle moduli in the monad map is at most $37(r+\tilde{r})-100$. For $r+\tilde{r}=5$, this is 85, so we still under-parametrize. For $r=\tilde{r}=3$, this is $122$ and we over-parametrize and the additional 4 moduli are removed by $h^1({\cal O}_X(b_1-c_1))=h^1({\cal O}_X(b_2-c_1))=2$. This means for $r=\tilde{r}=3$ (or larger), we will have made all 108 bundle moduli that enter the Pfaffian computation explicit in the monad map. 

Let us summarize the result of this subsection. For all values of $r, \tilde{r}\geq2$, the monad~\eqref{e4} describes the same bundle as the double extension in Eqs.~\eqref{eq:ExensionSequence}, \eqref{e6}. However, for $r+ \tilde{r} \leq5$ the monad  captures only a portion of the full moduli space of the extension. There are additional flat directions in the bundle moduli space which are not seen 
by the polynomial map $f$ in eq.~\eqref{e4}. However, if $r, \tilde{r} \geq 3$ the monad map realizes the full bundle moduli space of the  associated extensions bundle. In these cases the number of parameters in $f$ 
(modulo over parametrization due to symmetries discussed just above) precisely matches the number of flat directions of the extension bundle~\eqref{eq:ExensionSequence}, \eqref{e6}.

We will now consider the case $r= \tilde{r} =3$. In this case, the monad map explicitly realizes the same moduli space as the extension bundle and we must get the same answer for the Pfaffians. We will now show that it is, indeed, true.


\subsubsection*{Checking Beasley-Witten cancellation}


In order to find the Pfaffians for the first $\bP^1$ factor, we tensor the monad with $\mathcal{O}_X(-1,0,0,0,0,0)$ and restrict to the first $\mathbbm{P}^1$ to obtain
\begin{align}
0\xrightarrow{~}V_i\xrightarrow{~}\mathcal{O}_{\mathbbm{P}^1}\oplus\mathcal{O}_{\mathbbm{P}^1}\oplus\mathcal{O}_{\mathbbm{P}^1}(-1)\oplus\mathcal{O}_{\mathbbm{P}^1}(1)\xrightarrow{f|_{\gamma_i}}\mathcal{O}_{\mathbbm{P}^1}(3)\xrightarrow{~}0\,.
\end{align}
Next, we look at the sections $H^0(V_i)$. Since the line bundle $\mathcal{O}_{\mathbbm{P}^1}(-1)$ does not have any sections, we drop them from the subsequent discussion and compute
\begin{align}
H^0(V_i)\cong\text{Ker}[f|_{\gamma_i}:H^0(\mathcal{O}_{\mathbbm{P}^1}\oplus\mathcal{O}_{\mathbbm{P}^1}\oplus\mathcal{O}_{\mathbbm{P}^1}(1))\rightarrow H^0(\mathcal{O}_{\mathbbm{P}^1}(3)))]\,,
\end{align}
where $i=1,\ldots,8$ labels the eight curves in the curve class of the first $\mathbbm{P}^1$. 

In order to find the kernel of the map, let us parametrize the LHS of the map as 
$(\kappa_1, \kappa_2, z_{1,0}, z_{1,1})$ with $\kappa_1,\kappa_2\in\mathbbm{C}$ and $(z_{1,0}, z_{1,1})$ being a basis of holomorphic section of $\mathcal{O}_{\mathbbm{P}^1}(1)$.
Now we look at the restricted monad map after dropping the third columns that correspond to the $\mathcal{O}_{\mathbbm{P}^1}(-1)$ term,
\begin{align}
f(\vec{z}_{1};y)|_{\gamma_i}=(f^{(1)}_{(3, 0, 0, 0, 0, 0)}(\vec{z}_{1};y_i)\,,~f^{(2)}_{(3, 0, 0, 0, 0, 0)}(\vec{z}_{1};y_i)\,,~f^{(4)}_{(2, 5, 0, 0, 1, 2)}(\vec{z}_{1};y_i))\,.
\end{align}
Let us now study the conditions for which $(\kappa_1,\kappa_2,z_{2,0},z_{2,1})$ is in the kernel of the map. The first two components require
\begin{align}
f^{(1)}_{(3, 0, 0, 0, 0, 0)}(\vec{z}_{1};y)=0\,,\qquad f^{(2)}_{(3, 0, 0, 0, 0, 0)}(\vec{z}_{1};y)=0\, .
\end{align}
For the last component, we write
\begin{align}
\begin{split}
f^{(4)}_{(2, 5, 0, 0, 1, 2)}(\vec{z}_{1};y)&=z_{2,0}^2\tilde{f}^{(1)}_{(0, 5, 0, 0, 1, 2)}(y)+2z_{2,0}z_{2,1}\tilde{f}^{(2)}_{(0, 5, 0, 0, 1, 2)}(y)+z_{2,1}^2\tilde{f}^{(3)}_{(0, 5, 0, 0, 1, 2)}(y)\\
&=(z_{2,0},~z_{2,1})\cdot\begin{pmatrix}
\tilde{f}^{(1)}_{(0, 5, 0, 0, 1, 2)}(y) &\tilde{f}^{(2)}_{(0, 5, 0, 0, 1, 2)}(y)\\
\tilde{f}^{(2)}_{(0, 5, 0, 0, 1, 2)}(y) &\tilde{f}^{(3)}_{(0, 5, 0, 0, 1, 2)}(y)
\end{pmatrix}
\cdot
\begin{pmatrix}
z_{2,0}\\
z_{2,1}
\end{pmatrix}
\,.
\end{split}
\end{align}
This map has a non-trivial kernel iff
\begin{align}
\label{eq:Monadrrtilde3}
\det\left[
\begin{pmatrix}
\tilde{f}^{(1)}_{(0, 5, 0, 0, 1, 2)}(y) &\tilde{f}^{(2)}_{(0, 5, 0, 0, 1, 2)}(y)\\
\tilde{f}^{(2)}_{(0, 5, 0, 0, 1, 2)}(y) &\tilde{f}^{(3)}_{(0, 5, 0, 0, 1, 2)} (y)
\end{pmatrix}
\right]=0 \, .
\end{align}
This implies that the Pfaffian is proportional to the polynomial $\delta$ given by 
 \begin{align}
\label{e7}
\delta =\tilde{f}^{(1)}_{(0, 5, 0, 0, 1, 2)}(y)  \tilde{f}^{(3)}_{(0, 5, 0, 0, 1, 2)}(y)     -[\tilde{f}^{(2)}_{(0, 5, 0, 0, 1, 2)}(y) ]^2\,.
\end{align}
This is indeed the same expression as we have obtained for the extension bundle in Eq.~\eqref{eq:PfaffianExtensionBundle} and, hence, we have a non-trivial consistency check of our methods. We already know that the equation
\begin{align}
\sum_{i=1}^8 \lambda_i \delta_i=0
\end{align}
does not have solutions for $\lambda_i \in \mathbbm{C}^*$. Hence, a cancellation of the various Pfaffians cannot occur and the instanton superpotential $W_\gamma$ is non-zero.

 
\subsection{Discussion of the residue theorem}


We have constructed explicit models (and will indeed present large numbers of such models in a companion paper~\cite{Buchbinder:2019aaa}) which descend from the ambient space and appear to satisfy the conditions of the Beasley--Witten residue theorem, yet lead to non-vanishing instanton superpotentials. It is important to understand how this apparent contradiction is resolved. 

The first possibility is that a generic heterotic compactification involves five-branes. On the other hand it is not known whether 
the Beasley--Witten residue theorem is valid, or even applicable, in this case. However, in some cases we can find the hidden bundle, $V_\text{hid}$, which, together with the visible bundle $V$, satisfies the Bianchi identities without any need for five-branes. For example, the model discussed in subsection~\ref{sec:Extension_Example} admits the following hidden bundle: 
\begin{align}
V_\text{hid}= L_4\oplus L_5
\end{align}
with
\begin{align}
L_4=\cO_X(3, 2, -1, -1)\,,\qquad L_5=L_4^*=\cO_X(-3, -2, 1, 1)\,.
\end{align}
Indeed, one can check that $c_2 (TX)= c_2 (V) + c_2 (V_\text{hid})$ for this choice of hidden bundle. Nevertheless, a non-vanishing superpotential is found in this case. 

The only assumption of the residue theorem which we have not verified is  ``compactness of the instanton moduli space". Therefore, it is natural to propose that the models described in this section violate this condition. This assumption, however, is very elusive from a geometric viewpoint. It can only be checked in the framework of GLSMs where Bertolini and Plesser have established a precise criterion~\cite{Bertolini:2014dna}. In the remainder of this paper, we will, therefore, construct models with both a geometric and a GLSM description.


\section{Instantons in gauged linear sigma models}
\label{sec:GLSMs}


In this section, we will first review key facts about gauged linear sigma models. 
Then we will use the Bertolini-Plesser result~\cite{Bertolini:2014dna}, which provides a condition to check compactness of the instanton moduli space.


\subsection{Review of gauged linear sigma models}
\label{sec:GLSMReview}


Abelian gauged linear sigma models, or GLSMs for short, were introduced by Witten in Ref.~\cite{Witten:1993yc}. We will be mainly 
following the conventions and notation\footnote{To make the connection to the geometric approach more apparent, however, we will use the symbols $A$ and $B$ for the bundles.} 
of Ref.~\cite{Blaszczyk:2011hs}.  More details can also be found for example in Refs.~\cite{Distler:1992gi,Distler:1995mi}.

The representations of an $\mathcal{N}=(0,2)$ SUSY in two dimensions allow for chiral and vector superfields (as is familiar from the 4D $\mathcal{N}=1$ theory), 
but they can also have chiral-fermi and fermionic gauge fields, whose lowest components are fermions. Let us introduce a GLSM related  to an ambient space $\cP=\mathbb{P}^{n_1}\times\cdots \times\mathbb{P}^{n_m}$, a CICY described by a configuration matrix~\eqref{e10} and a two-term monad of the form~\eqref{eq:TwoTermMonad}. Its field content is as follows:
\begin{itemize}
\item A set of \U1 gauge superfields $(V,A)^{i}$, $i=1,\ldots, m$.
\item A set of fermionic gauge superfields $\Sigma^j$, $j=1,\ldots,n_F$, which, however, do not appear in two-term monads.
\item A set of chiral multiplets $\mathcal{Z}_I=(z_I,\psi_I)$, $I=1,\ldots N$ with $N=m+\sum_{i=1}^m n_i$ and charges $Q_I^i$ under the $i^\text{th}$ \U1.
\item A set of chiral multiplets $B_\beta=(p_\beta,\pi_\beta)$, $\beta=1,\ldots,r_B$ with charges $-b_\beta^i$ under the $i^\text{th}$ \U1.
\item A set of chiral-fermi multiplets $P_k=(c_k,\chi_k)$, $k=1,\ldots, K$ with charges $-q_k^i$ under the $i^\text{th}$ \U1.
\item A set of chiral-fermi multiplets $A_\alpha=(\lambda_\alpha, L_k)$, $\alpha=1,\ldots,r_A$ with charges $a_\alpha^i$ under the $i^\text{th}$ \U1.
\end{itemize}
Note that the names of the GLSM fields, the range of their indices and their charges are in line with the corresponding quantities from our geometric discussion. 

Each gauge superfield comes with FI-parameters $t_i$, $i=1,\ldots,m$ (and a $\theta$-angle), which are linked to the K\"ahler parameters of the geometric setup. Since we focus on favorable cases, the number of K\"ahler parameters is equal to the number of ambient space $\mathbbm{P}^{n_i}$ factors, and we have introduced that many \U1 gauge fields. The Calabi--Yau phase of the GLSM will correspond to the case where all $t_i\gg0$.

In addition to the gauge charges, all fields also carry a charge under an Abelian (non-$R$) symmetry $\U1_L$ and an Abelian $R$-symmetry $\U1_R$. In particular, the superpotential has $\U1_L$ charge 0 and $\U1_R$ charge~1. We summarize the charge assignments in Table~\ref{tab:GLSMChargeAssignments}.

\begin{table}[t]
\centering
\begin{tabular}{|c||c|c|c|c||c|c|}
\hline
				&$\mathcal{Z}_I$	&$P_k$			&$A_\alpha$	&$B_\beta$		&$S$		&$\Xi$			\\
\hline
$\U1^i$	&$Q_I^i$				&$-q_k^i$	&$a_\alpha^i$	&$-b_\beta^i$	&$Q_S$	&$Q_\Xi$		\\
$\U1_L$	&$0$					&$0$			&$-1$				&$1$				&$1$		&$-1$			\\
$\U1_R$	&$0$					&$1$			&$0$				&$1$				&$1$		&$0$			\\
\hline
\multirow{2}{*}{Interpretation}&Geometry&Geometry&Monad&Monad  &\multirow{2}{*}{Spectator}&\multirow{2}{*}{Spectator}\\
&coordinates&constraints&$A$-terms&$B$-terms&&\\
\hline
\end{tabular}
\caption{Overview of GLSM fields and their \U1 charges.}
\label{tab:GLSMChargeAssignments}
\end{table}

Note that in $(0,2)$ GLSMs the sum of the scalar charges can be non-zero, which will lead to a one-loop running of the FI parameters. In order to cure this, it was observed in~\cite{Distler:1995mi} that one can simply add a pair of spectator superfields, consisting of a chiral superfield $S$ and a chiral-fermi superfield $\Xi$ with opposite charges,
\begin{align}
Q_S^i=\sum_\beta b_\beta^i - \sum_k q_k^i\,,\qquad Q_\Xi^i=-\sum_\beta b_\beta^i + \sum_k q_k^i\,.
\label{e13}
\end{align}

The $\U1_{L,R}$ and the $\U1^i$ charges of $S$ and $\Xi$ allow for a term
\begin{align}
W\supset m\, S\Xi
\end{align}
in the superpotential so that the spectators pair up and become massive in the IR. However, as was shown in Ref.~\cite{Bertolini:2014dna}, the zero modes of $S$ can still decompactify the instanton moduli space.

Moreover, we note that the $\U1_{L,R}$ charge assignments are compatible with the superpotential
\begin{align}
W\supset \sum_{k=1}^K P_k H_k(\mathcal{Z}_I) + \sum_{\beta=1}^{r_B} B_\beta \sum_{\alpha=1}^{r_A} A_\alpha f_{\alpha\beta}(\mathcal{Z}_I)\,,
\end{align}
where $H_k(Z_I)$ and $f_\alpha(Z_I)$ are holomorphic polynomials in $\mathcal{Z}_I$ whose multi-degree is such that the superpotential is gauge-invariant. Note that $f$ corresponds to the monad map in Eq.~\eqref{eq:TwoTermMonad}.

In the Calabi--Yau phase $t_i\gg0$, the F- and D-terms give rise (for sufficiently generic $H$ and $f$ such that the geometry is smooth) to constraints
\begin{align}
H_k(\mathcal{Z}_I)\stackrel!=0\,, \qquad \forall k=1,\ldots,K\,,
\end{align}
which precisely imposes the $K$ equations that define the CICY in the ambient space. Since the $z_I$ are the collection of all $N$ coordinates of all $m$ ambient space factors, it makes sense to break them up into the $\mathbbm{P}^{n_i}$ they belong to, that is, to split $I$ into $I=\{i,r\}$ with $i=1,\ldots,m$ and $r=0,\ldots,n_i$. Then, the charge of $\mathcal{Z}_{i,r}^j$ under the $j^\text{th}$ \U1 is simply $\delta_i^j$ and $z_{i,0},\ldots,z_{i,n_i}$ are the homogeneous coordinates of $\mathbbm{P}^{n_i}$.

Thus, a GLSM as defined in Table~\ref{tab:GLSMChargeAssignments} describes a CICY given by the configuration matrix~\eqref{e10},
and a vector bundle $V$ on $X$ is given by a monad $0\to V\to A\xrightarrow{f} B\to 0$ with 
\begin{align}
\label{e12}
A=\bigoplus_{\alpha=1}^{r_A}\cO_X(a_\alpha^1,\ldots,a_\alpha^{m})\,,\qquad 
B=\bigoplus_{\beta=1}^{r_B}\cO_X(b_\beta^1,\ldots,b_\beta^{m})\,.
\end{align}

Since the GLSM is chiral, we have to worry about gauge anomalies. They lead to linear and quadratic constraints on the gauge charges. (Note that the contributions of the spectators to the anomalies cancel.) The linear anomaly constraints
\begin{align}
\sum_{I=1}^NQ_I^i=\sum_{k=1}^Kq_k^i\,, \qquad\sum_{\alpha=1}^{r_A} a_\alpha^i = \sum_{\beta=1}^{r_B} b_\beta^i\,,\qquad \forall\,i=1,\ldots,m\,,
\end{align}
precisely correspond to the conditions $c_1(TX)=0$ and $c_1(V)=0$, respectively. The quadratic conditions from the mixed $\U1^i\times\U1^j$ anomalies
\begin{align}
\label{eq:GLSMAnomaly}
0\stackrel!=A_{ij}=\left(\sum_{\alpha=1}^{r_A} a_\alpha^i a_\alpha^j - \sum_{\beta=1}^{r_B} b_\beta^i b_\beta^j \right) - \left(\sum_{I=1}^N Q_I^i Q_I^j - \sum_{k=1}^K q_k^i q_k^j \right)
\end{align}
are somewhat more mysterious. While it is true that $A_{ij}=0$ for all $i,j$ implies the Bianchi identities $c_2(V)=c_2(TX)$, the conditions are much stronger. Moreover, they depend on the chosen description of the geometry and the bundle, that is, different geometric descriptions of the same CY manifold can lead to different anomalies. Similarly, different descriptions of the same monad bundle on the same CY manifold can lead to different anomalies. 

Indeed, if we expand the $(2,2)$-forms $c_2(V)$ and $c_2(TX)$ in a basis of $(1,1)$-forms $D_i$,
\begin{align}
c_2(V) = \sum_{i,j}\nu_{i,j} D_i \wedge D_j\,,\qquad  c_2(TX) = \sum_{i,j}\gamma_{i,j} D_i \wedge D_j\,,
\end{align}
with $\gamma_{i,j}$ and $\nu_{i,j}$ in $\mathbbm{Z}$, we find that the quadratic anomaly coefficients $A_{i,j}$ appear in the Bianchi identity as
\begin{align}
c_2(V) -  c_2(TX) = \sum_{i,j} A_{ij} D_i \wedge D_j\,,\qquad A_{ij}= (\nu_{i,j}-\gamma_{i,j} )\,.
\end{align}
Hence, $A_{ij}=0$ is stronger than the Bianchi identities, since the (2,2)-forms $D_i \wedge D_j$, seen as forms on $X$, are not necessarily linearly independent in cohomology. (They are, however, linearly independent as forms on the ambient space.) Also, the GLSM anomaly conditions are independent of the GLSM phase one is considering.

To illustrate this further, let us look, for example, at a Calabi--Yau one-fold, that is a two-torus, with trivial bundle (no $A$ and $B$ fields) and a geometric description either as a cubic curve in $\mathbbm{P}^2$ or as a degree $6$ curve inside the weighted projective space $\mathbbm{P}_{1,2,3}$. In either case, there is only a single ambient space factor (and, hence, one \U1), a single constraint field $P$, and three coordinate fields $\mathcal{Z}_I$ with charges
\begin{align}
\mathbbm{P}^2[3]: Q_I=(1,1,1)\,,~-q_1=3\,,\qquad\qquad \mathbbm{P}_{1,2,3}[6]: Q_I=(1,2,3)\,,~-q_1=6\,.
\end{align}
Consequently, the anomaly $A_{11}$ in Eq.~\eqref{eq:GLSMAnomaly} is 6 and 12, respectively. From a geometric point of view, this is curious, since $c_2(TX)$ vanishes.  If we choose another description for the trivial bundle in terms of a three-term monad (which would correspond to the standard embedding)
\begin{align}
0\to\mathcal{O}\to \bigoplus_{\alpha=1}^3 A_\alpha \to B \to 0
\end{align}
with 
\begin{align}
\begin{split}
\mathbbm{P}^2[3]: &\quad a_\alpha=Q_I=(1,1,1)\,,~-b_1=-q_1=3\,,\\ 
\mathbbm{P}_{1,2,3}[6]: &\quad a_\alpha=Q_I=(1,2,3)\,,~-b_1=-q_1=6\,,
\end{split}
\end{align}
the anomalies vanish.

The observation that the GLSM anomalies seem to be too strong has motivated the introduction of a Green-Schwarz-like 
anomaly cancelation mechanism in the GLSM~\cite{Blaszczyk:2011ib,Quigley:2011pv}, which is generically necessary 
in the presence of flux that necessitates \SU3 structure compactifications~\cite{Adams:2006kb,Adams:2009av}. The mechanism introduces 
field-dependent FI terms, and the gauge transformation of the fields shift the path integral measure to cancel the anomalies. However, since the field-dependent FI terms 
correct the K\"ahler parameters, the K\"ahler form will no longer be closed and the CY manifold is deformed to a geometry with torsion
(in case there is an anomaly despite $c_2(V)=c_2(TX)$) or with five-branes that cancel the geometric anomaly (in cases where $c_2(V)<c_2(TX)$). It would be interesting to study the consequences for the compactness of the instanton moduli space and, consequently, for the Beasley-Witten theorem in this context. This is, however, beyond the scope of this paper. 

\subsection{Compactness of the instanton moduli space for monad bundles}
The conditions for a compact instanton moduli space have been identified by Bertolini and Plesser in Ref.~\cite{Bertolini:2014dna}. They found 
that the zero modes of the bosonic chiral multiplets $B$ and $S$ decompactify the instanton moduli space. 
Let us expand an effective curve $\mu$ in a curve class $\gamma$ in terms of Mori cone generators  $\mu_i$ such that $\mu=\sum_i w_i \mu_i$ with $w_i\geq0$. The instantons in this curve class will have instanton numbers $(w_1,w_2,\ldots,w_m)$. 

The decompactifying zero modes are in one-to-one correspondence with the lowest component fields $p_{\beta}$ and $s$ of the superfields 
$B_\beta$ and $S$. On $\mathbbm{P}^1$, these
zero modes are given by the sections of the corresponding line bundles twisted with the $\mathbbm{P}^1$ 
spin bundle $K^{1/2}(\mathbbm{P}^1)=\cO_{\mathbbm{P}^1}(-1)$, 
\begin{align}
\label{eq:BP_Conditions1}
\begin{split}
p_\beta 	&~~\leftrightarrow~~ \Gamma(\mathcal{O}_{\mathbbm{P}^1}(-b_\beta^i w_i-1))\,,\\
s				&~~\leftrightarrow~~ \Gamma(\mathcal{O}_{\mathbbm{P}^1}(\;\,Q_S^i w_i-1))\,.
\end{split}
\end{align}
Note that if $s$ has zero modes outside the geometric cone, the instanton moduli space will nevertheless be compact. Hence, by using Bott's formula we find that there will be decompactifying sections if for any $\beta$ there exist $w_i$ such that 
\begin{align}
\label{eq:BP_Conditions2}
-b_\beta^i w_i-1\geq 0\quad\text{or}\quad Q_S^i w_i-1\geq0\,. 
\end{align}

\subsection{Relation to the geometric models}

In the next section, we will be interested in geometric heterotic models of the kind studied previously that admit a GLSM description. Let us discuss a special case of Eq.~\eqref{eq:BP_Conditions2}, where the instanton wraps a single genus zero curve once so that $(w_1,w_2,\ldots ,w_m)=(1,0,\ldots,0)$. Moreover, we consider semi-positive monads which 
satisfy $b_\beta^i \geq0$ since  they have a better chance of giving rise to poly-stable vector bundles. 
This, implies that the first condition in~\eqref{eq:BP_Conditions2} is not satisfied and 
we will never find decompactifying zero modes from the monad fields $p_\beta$. However, the zero modes of the spectator $s$ can still lead to a non-compact instanton moduli space. Since we focus on a single ambient space $\mathbbm{P}^1$,
the second equation in~\eqref{eq:BP_Conditions2} becomes 
\begin{align}
Q_S^1 \geq 1\,. 
\label{e12.0}
\end{align}
Using Eq.~\eqref{e13} and the fact that $\sum_k q_k^2 =2$ (which is just the Calabi--Yau condition) we can equivalently write this as 
\begin{align}
\label{eq:SimpleMonadBWViolated}
\sum_\beta b_\beta^1\geq3\,.
\end{align}
It would be very interesting to find a reason why monads of the type we consider with $\sum_\beta b_\beta^1<3$ always have a compact instanton moduli space. 

A second very interesting question is the interplay of anomalies with the compactness criterion. The compactness criterion is formulated in terms of monad fields $B_\beta$ on the $\mathbbm{P}^1$, while the geometric calculations mainly depend on  the transverse space $\mathcal{Q}$. However, the anomaly conditions mixes data for $\mathbb{P}^1$ with that for $\mathcal{Q}$, so the bundle charges in the $\mathbbm{P}^1$ direction implicitly depend on all other charges as well. It would be interesting to use the anomalies to establish this connection explicitly. While we do not do this in full generality, we illustrate how the anomalies constrain the bundle charges for a simple class of monads in Section~\ref{sec:SimpleMonadAnomalyFree}. 

When the GLSM anomalies are not cancelled the model is not consistent and we cannot invoke the Bertolini-Plesser criterion to check compactness of the moduli space.
Indeed, if we check compactness naively for such anomalous GLSMs, we find cases with an (apparently) compact moduli space but Pfaffians computed from geometric methods which are linearly independent. For such cases we would expect that, upon constructing an equivalent but anomaly-free description, the fields $B'_\beta$ of the new model will lead to zero modes (they will either have zero modes themselves or their charges will change the spectator superfields such that $S$ develops zero modes) that decompactify the instanton moduli space.

\subsection{Example for simple monad bundles}
\label{sec:SimpleMonadAnomalyFree}


Let us illustrate how the requirement of vanishing GLSM anomalies constrains the possible bundle charges in some simple cases. First, we consider an \SU3 monad bundle with just one $B$ term,
\begin{align}
0\to V \to \bigoplus_{\alpha=1}^4 \cO_X(a_\alpha,\hat{a}_\alpha) \to \cO_X(b,\hat{b}) \to 0\,,
\label{e14}
\end{align}
where $a_\alpha$ and $b$ denote the degrees in the $\mathbb{P}^1$ direction and $\hat{a}_\alpha$  and $\hat{b}$ are the multi-degrees in the transverse space $\mathcal{Q}$. We focus on semi-positive monads where $a_\alpha, \hat{a}_\alpha$ (and consequently $b$ and $\hat{b}$ as well) 
are non-negative.

A vanishing first Chern class of $V$ 
requires 
\begin{align}
\sum_{\alpha=1}^4 a_\alpha = b\,,\qquad \sum_{\alpha=1}^4 \hat{a}_\alpha = \hat{b}\,.
\label{e20}
\end{align}
The quadratic anomaly of the first \U1, that is,\ $A_{11}$ in~\eqref{eq:GLSMAnomaly}, then imposes
\begin{align}
\label{eq:QuadraticAnomalyFirstFactor}
\text{type I:}~~\sum_{\alpha<\alpha'} a_\alpha a_{\alpha'} =0\,,\qquad
\text{type II:}~~\sum_{\alpha<\alpha'} a_\alpha a_{\alpha'} =1\,,
\end{align}
where we have also used Eq.~\eqref{e20}. Since we only consider monads with $a_\alpha\geq0$, we find that at least two $a_\alpha$ have to be zero, and we take, without loss of generality, $a_3=a_4=0$. Now, the type I cases require that in addition we choose (say) $a_2=0$ in which case the quadratic anomaly $A_{11}$ vanishes identically for any $a_1$. In particular, if $a_1\leq2$, the Beasley-Witten vanishing conditions will be satisfied from Eq.~\eqref{eq:SimpleMonadBWViolated} (since $b=a_1$), while $a_1>2$ will lead to a non-compact instanton moduli space on this $\mathbbm{P}^1$. For the type II cases, the solutions are $a_3=a_4=0$ and $a_1=a_2=1$. Hence, these always lead to compact instanton moduli spaces. 

While the problem of solving the quadratic Diophantine equations corresponding to the mixed anomalies is too general to make further precise statements, it is instructive to look at the mixed quadratic anomalies $A_{1i}$ with $i>1$. Here we have to distinguish the type I and type II cases.

\subsubsection*{Type I case}
The mixed quadratic anomalies $A_{1i}$ for type I read
\begin{align}
\label{eq:MixedAnomalyType1}
q_1^i+q_2^i = a_1 (b^i-a_1^i)\,.
\end{align}
Let us observe that the vector $\hat{b}- \hat{a}_1= (b^i-a_1^i)$ on the RHS gives precisely the multi-degree of the monad map $f$ in the $i^\text{th}$ ambient 
space factor coordinates. In the next section, we will see that only the first column of the monad map 
will contribute to the Pfaffian. Hence, this will determine the multi-degrees of the Pfaffian in the coordinates of the projective ambient space.

Similarly, the two terms $q_1^i$ and $q_2^i$ are the degrees of the defining equation in the $i^\text{th}$ ambient space factor. 
These multiply the $\mathbbm{P}^1$ coordinates and, hence, enter in the Gromov-Witten invariants, see Eq.~\eqref{eq:CICYCurves}. This means that the degree of the Pfaffian is fixed entirely in terms of the twisting of the $\bP^1$ over $\mathcal{Q}$.

While it is hard to make a definite statement, we observe that the following relation 
\begin{align}
a_1\leq q_1^i+q_2^i\leq4
\end{align}
gives a necessary condition for vanishing of the $A_{1i}$ anomaly. Here, the final bound comes from the fact that the highest $\bP^1$ twisting that occurs in the CICY list is~$4$. Since for type I we have $a_1= b$, we conclude, using eq.~\eqref{eq:SimpleMonadBWViolated}, that the window of possible \SU3 monads with a single $B$ field that have a non-compact instanton moduli space is very narrow.

\subsubsection*{Type II case}
For type II, the mixed anomalies $A_{1i}$ become
\begin{align}
\label{eq:MixedAnomalyType2}
2q_1^i = 2 b^i- a_1^i-  a_2^i\,.
\end{align}
Again, we will see in the next section that  the RHS is precisely the degree of the Pfaffian
and this degree is fixed for a given CICY by the twisting of the $\bP^1$ over $\mathcal{Q}$. Similar to the first case, we find a  necessary condition 
for the mixed anomalies $A_{1i}$ to vanish
\begin{align}
b^i\leq q_1^i\leq4\,.
\end{align}
In this case, the Pfaffian will be given by the determinant of a matrix, which will lead to a polynomial of multi-degree $2 b^i- a_1^i-  a_2^i$.



\section{Geometric models with a GLSM description}
\label{sec:Compactness}

In this section, we will discuss heterotic models for which we can find a GLSM description. First, in Section~\ref{sec:PfaffiansSU3General}, we will discuss in general the functional form of the Pfaffians that we will encounter in the examples of this section. Then, in Section~\ref{sec:Monad_Examples_Vanishing_TypeIAndII}, we present examples with a compact instanton moduli space. For these models, the superpotential must vanishes according to the the results of Refs.~\cite{Silverstein:1995re, Basu:2003bq, Beasley:2003fx}. Indeed, we will see that in these cases the Pfaffians computed from geometric methods are always linearly dependent and, hence, allow for a cancellation. This provides a non-trivial consistency check between the geometric methods and the vanishing theorems. Finally, in Section~\ref{sec:Monad_Examples_Vanishing_MoreComplicated}, we discuss SU(3) monads for which the spectator decompactifies the instanton moduli space.

In our scan we also find about 50 models with an anomaly-free GLSM description where the Bertolini--Plesser criterion is violated and the instanton moduli space is, hence, non-compact. Nevertheless, in all such cases the Pfaffians computed via geometric methods turn out to be linearly dependent. This is surprising and counter to expectations but it is not a contradiction. Indeed, for such cases neither the GLSM nor the geometric approach tell us whether the instanton superpotential is vanishing or non-vanishing. 

A non-vanishing instanton superpotential for these cases is consistent with the geometric picture -- the values of the undetermined constants $\lambda_i$ may be such that the linearly dependent Pfaffians do not cancel -- and would point to our geometric non-vanishing condition being too weak. On the other hand, vanishing of the instanton superpotential for these models -- perhaps the more plausible case since the emergence of linearly dependent Pfaffians is a highly non-trivial feature -- would hint at reasons for the cancellation of instanton effects which go beyond the current vanishing theorems.  It would clearly be important to understand which of these possibilities is realized. For now, we emphasize that in all cases where both a geometric and an anomaly-free GLSM description exist, there is no contradiction between the geometric results and the vanishing theorems.

\subsection{Computing Pfaffians for SU(3) monad bundles}
\label{sec:PfaffiansSU3General}
Our examples with compact instanton moduli space in this section are based on the SU(3) monad bundles 
introduced in subsection~\ref{sec:SimpleMonadAnomalyFree}.
Let us first discuss the computation in some generality, and then present an example for each type. 
For type~I examples, the monad~\eqref{e14} restricted to the $\mathbbm{P}^1$ and twisted by the spin bundle becomes
\begin{align}
\label{eq:TypeISimpleExample}
0\to V|_{\mathbbm{P}^1}\otimes\cO_{\mathbbm{P^1}}(-1)\to 
\cO_{\mathbbm{P^1}}(a_1-1)\oplus \cO_{\mathbbm{P^1}}(-1)\oplus \cO_{\mathbbm{P^1}}(-1)
\oplus \cO_{\mathbbm{P^1}}(-1)\to \cO_{\mathbbm{P^1}}(a_1-1)\to 0\,,
\end{align}
where we used $b=a_1$ as well other consequences of the anomaly cancellation from subsection~\ref{sec:SimpleMonadAnomalyFree}.

Since the Pfaffian will be proportional to the locus where the number of sections jumps, we can discard the $\cO_{\mathbbm{P^1}}(-1)$ factors which have no sections and focus on the monad map $f^{(1)}(y)$ with $y=(\vec{z}_2,\ldots,\vec{z}_{n_m})$ from ${\cal O}_X(a_1)$ to $B$. The zero locus of this map will be proportional to the Pfaffians
\begin{align}
\label{eq:MonadTypeIGeneral}
\delta=f_{\hat{b}-\hat{a}_1}^{(1)}(y)\,,
\end{align}
where the subscript denotes the degrees in the coordinates $y=(\vec{z}_2,\ldots,\vec{z}_{n_m})$ of $\mathcal{Q}$. Note that, 
as mentioned in subsection~\ref{sec:SimpleMonadAnomalyFree}, the multi-degree of the Pfaffian is 
precisely the RHS of the mixed anomaly condition~\eqref{eq:MixedAnomalyType1}. The coefficient of each monomial in $\delta$ corresponds to a bundle modulus. 
As discussed earlier, these might over- or under-parametrize  the bundle moduli space. Next, we find the position of the curves in the transverse space $\mathcal{Q}$, insert these into~\eqref{eq:MonadTypeIGeneral} and compute $\delta$ for each curve. Then we check whether or not the $\delta_i$ are linearly independent. As we already discussed in subsection~\ref{sec:SimpleMonadAnomalyFree}, the compactness conditions are satisfied for a non-anomalous GLSM and a bundle with $a_1\leq2$ and, hence, $\delta_i$ must be linearly dependent. We will present 
an explicit example below.

Similarly, for the type II examples, the monad~\eqref{e14} restricted to the $\mathbbm{P}^1$ and twisted by the spin bundle becomes
\begin{align}
\label{eq:TypeIISimpleExample}
0\to V|_{\mathbbm{P}^1}\otimes\cO_{\mathbbm{P^1}}(-1)\to \cO_{\mathbbm{P^1}}(1)\oplus 
\cO_{\mathbbm{P^1}}(1)\oplus \cO_{\mathbbm{P^1}}(-1)\oplus \cO_{\mathbbm{P^1}}(-1)\to \cO_{\mathbbm{P^1}}(2)\to 0\,.
\end{align}
We can again drop the $\cO_{\mathbbm{P^1}}(-1)$ parts and focus on the first two columns $f^{(1)}$ and $f^{(2)}$ of the monad map. Choosing a basis $(z_{1,0}, z_{1,1})$ for the sections of $\cO_{\mathbbm{P^1}}(1)$ and $(z_{1,0}^2,z_{1,0} z_{1,1}, z_{1,1}^2)$ for the sections of $\cO_{\mathbbm{P^1}}(2)$, we find for the maps $f^{(\alpha)}$ from ${\cal O}_X(a_\alpha)$ to $B$ (with $\alpha={1,2}$)
\begin{align}
f^{(\alpha)}=z_{1,0} \tilde{f}^{(\alpha,0)}(y) + z_{1,1}\tilde{f}^{(\alpha,1)}(y)
\end{align}
with $y=(\vec{z}_2,\ldots,\vec{z}_{n_m})$ and, hence, the map reads
\begin{align}
(z_{1,0},~z_{1,1})\cdot
\begin{pmatrix}
\tilde{f}^{(1,0)}_{\hat{b}-\hat{a}_1}(y) & \tilde{f}^{(2,0)}_{\hat{b}-\hat{a}_2}(y)\\[8pt]
\tilde{f}^{(1,1)}_{\hat{b}-\hat{a}_1}(y) & \tilde{f}^{(2,1)}_{\hat{b}-\hat{a}_2}(y)
\end{pmatrix}\cdot
\begin{pmatrix}z_{1,0}\\z_{1,1}\end{pmatrix}\,.
\end{align}
This has a non-trivial kernel if the determinant
\begin{align}
\label{eq:MonadTypeIIGeneral}
\delta&=\det\left[
\begin{pmatrix}
\tilde{f}^{(1,0)}_{\hat{b}-\hat{a}_1}(y) & \tilde{f}^{(2,0)}_{\hat{b}-\hat{a}_2}(y)\\[8pt]
\tilde{f}^{(1,1)}_{\hat{b}-\hat{a}_1}(y) & \tilde{f}^{(2,1)}_{\hat{b}-\hat{a}_2}(y)
\end{pmatrix}
\right]\\[4pt]
&=\tilde{\tilde{f}}^{(0)}_{2\hat{b}-\hat{a}_1-\hat{a}_1}(y)-\tilde{\tilde{f}}^{(1)}_{2\hat{b}-\hat{a}_1-\hat{a}_1}(y)
\end{align}
with
\begin{align}
\tilde{\tilde{f}}^{(0)}_{2\hat{b}-\hat{a}_1-\hat{a}_1}(y)=\tilde{f}^{(1,0)}_{\hat{b}-\hat{a}_1}(y) \tilde{f}^{(2,1)}_{\hat{b}-\hat{a}_2}(y)\,,\\[4pt]
\tilde{\tilde{f}}^{(1)}_{2\hat{b}-\hat{a}_1-\hat{a}_1}(y)=\tilde{f}^{(2,0)}_{\hat{b}-\hat{a}_2}(y)\tilde{f}^{(1,1)}_{\hat{b}-\hat{a}_1}(y) \,,
\end{align}
 vanishes.
Again, the degree of the Pfaffian is precisely the RHS of the mixed anomaly condition~\eqref{eq:MixedAnomalyType2} 
(cf.\ subsection~\ref{sec:SimpleMonadAnomalyFree}), and the coefficients of the monomials of the 
$\tilde{f}^{(\alpha,r)}$, could over- or under-parametrize the bundle moduli space. As the next step, we find the position of the curves in $\mathcal{Q}$, insert these into~\eqref{eq:MonadTypeIIGeneral} and compute $\delta$ for each curve. As was already explained in subsection~\ref{sec:SimpleMonadAnomalyFree}, in all type II models of this kind the compactness criteria are satisfied and, hence, $\delta_i$ must be linearly dependent. As a consistency check, we have verified, by scanning all 123 CICYs with Picard rank 3, that the geometric method always produces linearly dependent Pfaffians for bundles that satisfy the GLSM anomaly conditions. An explicit type II example is presented below.

Let us also give a setup where the compactness criterion fails. 
As was discussed in Eq.~\eqref{eq:SimpleMonadBWViolated}, we need monad bundles whose line bundles ${\cal O}_X(b_\alpha)$ sum up to at least three to have a 
spectator zero mode that decompactifies the instanton moduli space. Sticking with \SU3 bundles, we choose a construction with a rank five line bundle sum $A$ and a rank two line bundle sum $B$, explicitly given by
\begin{align}
A=\cO_X(2, \hat{a}_1)\oplus\cO_X(1, \hat{a}_2)\oplus\cO_X(0, \hat{a}_3)\oplus\cO_X(0, \hat{a}_4)\oplus\cO_X(0, \hat{a}_5),\;\;
B=\cO_X(2, \hat{b}_1)\oplus\cO_X(1, \hat{b}_2)\,.
\label{e30}
\end{align}
 Projecting to the $\mathbbm{P}^1$ factor and tensoring with $\cO_{\mathbbm{P}^1}(-1)$ gives
\begin{align}
\label{eq:MonadMoreComplicatedGeneral}
0\to V|_{\mathbbm{P}^1}\otimes\cO_{\mathbbm{P}^1}(-1)\to\cO_{\mathbbm{P}^1}(1)\oplus\cO_{\mathbbm{P}^1}(0)\oplus 3\cO_{\mathbbm{P}^1}(-1)\to \cO_{\mathbbm{P}^1}(1)\oplus\cO_{\mathbbm{P}^1}(0)\to0 \,.
\end{align}
As before, we drop the $\cO_{\mathbbm{P}^1}(-1)$ that do not have sections, so that the relevant part of the $5\times2$ monad map $f$ is the first $2\times2$ block,
\begin{align}
f_{2\times2}=
\begin{pmatrix}
f^{(1,1)}_{\hat{b}_1-\hat{a}_1}	& 0\\
f^{(2,1)}_{\hat{b}_1-\hat{a}_2}	& f^{(2,2)}_{\hat{b}_2-\hat{a}_2}
\end{pmatrix}\,,
\qquad \text{with}~ f^{(i,j)}\equiv 0 \text{~if~}\min(\hat{b}_j-\hat{a}_i)<0\,.
\end{align} 
Choosing a basis of sections of $\cO_{\mathbbm{P}^1}(1)$ as $(z_{1,0},z_{1,1})$ and of $\cO_{\mathbbm{P}^1}(0)$ as $\kappa\in\mathbbm{C}$, the map $\delta$ becomes

\begin{align}
\delta=
\det\left[
\begin{pmatrix}
\tilde{f}^{(1,1)}_{\hat{b}_1-\hat{a}_1}	& 0																					& \tilde{f}^{(2,1)}_{\hat{b}_1-\hat{a}_2}\\[8pt]
0															& \tilde{\tilde{f}}^{(1,1)}_{\hat{b}_1-\hat{a}_1}					& \tilde{\tilde{f}}^{(2,1)}_{\hat{b}_1-\hat{a}_2}\\[8pt]
0															& 0																					& f^{(2,2)}_{\hat{b}_2-\hat{a}_2}\\
\end{pmatrix}
\right]=\tilde{f}^{(1,1)}_{\hat{b}_1-\hat{a}_1}\tilde{\tilde{f}}^{(1,1)}_{\hat{b}_1-\hat{a}_1}f^{(2,2)}_{\hat{b}_2-\hat{a}_2}\,.
\end{align} 
The functions with tildes are functions of the same degree as the corresponding functions without tilde and their coefficients parametrize the moduli space.

Scanning over a large number of CICYs with Picard rank 3, we find the somewhat surprising result that all models with monads of the type above have linearly dependent Pfaffians. As discussed earlier, these results are consistent with either vanishing or non-vanishing of the instanton sum and it would be important to decide between these two possibilities. However, our experience points to an unexpected vanishing of the instanton superpotential which goes beyond the vanishing predictions of the Beasley-Witten theorem: more specifically, our examples typically involve $\mathcal{O}(10)$ curves in the class $\gamma$ under consideration, and Pfaffians which contain many thousands of different monomials in bundle moduli. The fact that these thousands of terms can sum to zero by choosing just $\mathcal{O}(10)$ 
constants $\lambda_i$ is extremely surprising and unlikely to be a coincidence. 

\subsection{Examples with compact moduli space}
\label{sec:Monad_Examples_Vanishing_TypeIAndII}

\subsubsection{The geometric computation}
We consider CICY 7836 with configuration matrix
\begin{align}
\label{eq:MonadExampleCICYConf7836}
X\in\left.\left[
\begin{array}{c|ccc}
\mathbbm{P}^{1}&1&1&0\\
\mathbbm{P}^{1}&2&0&0\\
\mathbbm{P}^{4}&2&1&2\\
\end{array}
\right]^{(3,61)}\quad \right| \quad
\begin{array}{l}
\vec{z}_1=[z_{1,0}:z_{1,1}]\\
\vec{z}_2=[z_{2,0}:z_{2,1}]\\
\vec{z}_3=[z_{3,0}:z_{3,1}:z_{3,2}:z_{3,3}:z_{3,4}]\,.
\end{array}
\end{align}
and the two-term monad $0\to V \to A \stackrel{f}{\to} B \to 0$ with
\begin{align}
A=\cO_X(1,0,0)\oplus\cO_X(0,1,0)\oplus\cO_X(0,0,1)\oplus\cO_X(0,1,2)\,,\quad B=\cO_X(1,2,3)\,.
\end{align}
This example allows us to discuss both the type I and the type II case by focusing on the first and second ambient space $\mathbbm{P}^1$ factor, respectively. By explicitly constructing the curves from Eq.~\eqref{eq:CurvesInBothCases}, we find that there are 16 curves in the curve class of the first $\bP^1$ and 40 curves in the curve class of the second $\bP^1$, respectively. This example satisfies all consistency conditions we need to impose on a consistent string compactification.

Let us also compute the dimension of the bundle moduli space. On the ambient space, the monad map $f$ is given by the $1\times 4$ matrix
\begin{align}
f=\left(f_{(0,2,3)}\,,\; f_{(1,1,3)}\,,\; f_{(1,2,2)}\,,\; f_{(1,1,1)}\right)\,.
\end{align}
By counting the number of monomials that appear in $f$, we find a total of $105+140+90+20=355$ terms. This (vastly) over-parametrizes the bundle 
moduli space, since $h^{1}(V^* \otimes V)=228$, as we shall demonstrate next. 

First, we observe that $h^1(B^*\otimes A)=h^1(B^*\otimes B)=0$, so that Eq.~\eqref{eq:MonadBundleEndDim} can be applied. In addition, $h^1(B^*\otimes A)=h^1(A^*\otimes A)=h^{0}(B^*\otimes A)=0$ but we have the contributions
\begin{align}
h^0({\cal O}_X(b_1-a_\alpha))=(89,92,68,18)\,.
\end{align}
This is to be contrasted with the number of moduli that appear in the ambient space monad map. Let us illustrate the reduction for the first term. Out of the original 105, 16 restrict to zero on $X$. This can be seen from the Koszul sequence together with
\begin{align}
h^0(\cN_{3}^*\otimes {\cal O}_\cP(b_1-a_1) )&=h^0(\cO_\cP(0,2,1))=15\\
 h^1(\cN_{1}^*\otimes \cN_{2}^*\otimes{\cal O}_\cP(b_1-a_1) )&=h^1(\cO_\cP(-2,0,0))=1\,.
\end{align}
Hence these $15+1$ contributions vanish on the CY $X$. The computations that illustrate the reduction for the cohomologies $H^0({\cal O}_X(b_1-a_\alpha))$ where $\alpha=2,3,4$ are analogous.

In addition to the above, we also have to subtract the degrees of freedom that can be attributed to the endomorphisms of the line bundle sums $A$ and $B$. Obviously, $h^0(B\otimes B^*)=1$. For the contributions to $h^0(A\otimes A^*)$, we find
\begin{align}
h^0({\cal O}_X(a_\alpha- a_{\alpha'}))=\begin{pmatrix}
1&0&1&10\\
0&1&0&14\\
0&0&1&10\\
0&0&0&1
\end{pmatrix}\,.
\end{align}
Combining everything, we arrive at the aforementioned result $h^{1}(V^* \otimes V)=228$.

\subsubsection*{Type I case}
The restriction of the monad bundle to the first $\bP^1$ corresponds to the case discussed in~\eqref{eq:TypeISimpleExample} with $a_1=1$. The Pfaffian is hence given by
\begin{align}
\delta=f^{(1)}_{2,3}(\vec{z}_2,\vec{z}_3)=\sum_{i=1}^{105}\beta_i m_i(y)\,.
\end{align}
This has 105 monomials $m_i$ and bundle moduli $\beta_i$ (89 of which are independent on $X$). While we could identify the 16 terms that restrict trivially to $X$, we do not need to in the computation of the Pfaffians. We take the position $y_i$ of the curves $\gamma_i$, $i=1,\ldots,16$, for an arbitrary but fixed position in complex structure moduli space, and insert them into $\delta$. Beasley-Witten cancellation then means that the sum
\begin{align}
\sum_{i=1}^{16}\lambda_i \delta(y_i)
\end{align}
can vanish by choosing the 16 $\lambda_i\in\mathbbm{C}^*$ appropriately. Indeed, we find that out of the 105 terms (with generic bundle moduli $a_i$), only 15 are linearly independent and, hence, the sum can cancel by choosing the $16$ coefficient $\lambda_i$ appropriately. 

\subsubsection*{Type II case}
Considering the contribution from the second $\mathbbm{P}^1$ instead, the restriction in this example corresponds to the one discussed in~\eqref{eq:TypeIISimpleExample}. The Pfaffian is hence given by
\begin{align}
\delta=\text{det}\left[
\begin{pmatrix}
\tilde{f}^{(1,0)}_{1,3}(y)&\tilde{f}^{(2,0)}_{1,1}(y)\\
\tilde{f}^{(1,1)}_{1,3}(y)&\tilde{f}^{(2,1)}_{1,1}(y)
\end{pmatrix}
\right]\,,
\end{align}
with $y=\vec{z}_1,\vec{z}_3$ and 
\begin{align}
\begin{split}
\tilde{f}^{(1,0)}_{1,3}(y)=\sum_{i=1}^{70} \beta_i^{(1,0)} m_{1,3}^i(y)\,,\qquad \tilde{f}^{(1,1)}_{1,3}(y)=\sum_{i=1}^{70} \beta_i^{(1,1)} m_{1,3}^i(y)\,,\\
\tilde{f}^{(2,0)}_{1,1}(y)=\sum_{i=1}^{10} \beta^{(2,0)}_i m_{1,1}^i(y)\,,\qquad \tilde{f}^{(2,1)}_{1,1}(y)=\sum_{i=1}^{10} \beta^{(2,1)}_i m_{1,1}^i(y)\,,\\
\end{split}\,,
\end{align}
where $m_{d_1,d_2}^i$ is the $i^\text{th}$ monomial of multi-degree $(d_1,d_2)$ in $y=(\vec{z}_1,\vec{z}_3)$. This determinant has 1400 terms bilinear in the moduli $\beta$.  Proceeding as before, we find the 40 curves at arbitrary but fixed complex structure and consider the sum
\begin{align}
\sum_{i=1}^{40}\lambda_i \delta(y_i)\,.
\end{align}
We find that out of the 1400 terms, 39 are linearly independent, so there exist $\lambda_i$, $i=1,\ldots,40$ such that the sum vanishes.

\subsubsection{The GLSM computation}
The corresponding GLSM can be easily constructed from this geometric data. We summarize the GLSM charges in Table~\ref{tab:GLSMChargesVanishingExamples}. The two cases fit into our general discussion in Section~\ref{sec:SimpleMonadAnomalyFree}, so we already know that the Bertolini-Plesser criteria are fulfilled and the 
instanton moduli space is compact.

\begin{table}[t]
\centering
\begin{tabular}{|c||c@{~~}c@{~~}c@{~~}|c@{~~}c@{~~}c@{~~}||c@{~~}c@{~~}c@{~~}c@{~~}|c@{~~}||c@{~~}|c@{~~}|}
\hline
&$\cZ_{1,2}$ & $\cZ_{3,4}$ & $\cZ_{5,6,7,8,9}$ & $P_1$  & $P_2$ & $P_3$ &$A_1$ & $A_2$ & $A_3$ & $A_4$ & $B$&$S$&$\Xi$\\
\hline
$\U1_1$ & 1&0&0  &-1&-1&0  	& 1&0&0&0		&-1		&-1 		& 1	\\
$\U1_2$ & 0&1&0  &-2&0&0		& 0&1&0&1		&-2  	&0		&0	\\
$\U1_3$ & 0&0&1  &-2&-1&-2	& 0&0&1&2		&-3  	&-2		&2	\\
\hline
\end{tabular}\newline
\caption{GLSM charges of the monad bundle on CICY 7836.}
\label{tab:GLSMChargesVanishingExamples}
\end{table}

Using~\eqref{eq:GLSMAnomaly}, we find that all GLSM anomalies cancel. Hence, we have an anomaly free GLSM description of 
this heterotic model. Using Eq.~\eqref{eq:BP_Conditions2}, we see from Table~\ref{tab:GLSMChargesVanishingExamples}
that neither the spectator field $S$ nor the field $B$ have zero modes. 
Hence, the GLSM model satisfies the compactness criterion by Bertolini and Plesser and the superpotential must be zero from the vanishing theorems~\cite{Silverstein:1995re, Basu:2003bq, Beasley:2003fx}. This nicely explains the seemingly miraculous cancelation of the 105 and 1400 terms in the Pfaffian for the two cases discussed above. It also provides a highly non-trivial consistency check of the geometric technique. 

\subsection{Example with non-compact moduli space}
\label{sec:Monad_Examples_Vanishing_MoreComplicated}


Let us also give an example which has an (anomaly free) GLSM description with non-compact moduli space. 
For this we consider CICY 7555, which is very similar to CICY 7834 discussed above. Its configuration matrix reads
\begin{align}
\label{eq:MonadExampleCICYConf7555}
X\in\left.\left[
\begin{array}{c|ccc}
\mathbbm{P}^{1}&1&1&0\\
\mathbbm{P}^{1}&2&0&0\\
\mathbbm{P}^{4}&1&2&2\\
\end{array}
\right]^{(3,61)}\quad \right| \quad
\begin{array}{l}
\vec{z}_1=[z_{1,0}:z_{1,1}]\\
\vec{z}_2=[z_{2,0}:z_{2,1}]\\
\vec{z}_3=[z_{3,0}:z_{3,1}:z_{3,2}:z_{3,3}:z_{3,4}]\,.\\
\end{array}
\end{align}
We furthermore consider the two-term monad $0\to V \to A \stackrel{f}{\to} B \to 0$ with
\begin{align}
\begin{split}
A&=\cO_X(2,1,0)\oplus\cO_X(1,0,2)\oplus\cO_X(0,1,0)\oplus\cO_X(0,0,1)\oplus\cO_X(0,0,1)\,,\\ 
B&=\cO_X(2,2,1)\oplus\cO_X(1,0,3)\,.
\end{split}
\end{align}
Again, this example satisfies all consistency conditions.

\subsubsection{The geometric computation}
The restriction of the monad bundle to the first $\bP^1$ has been discussed in \eqref{eq:MonadMoreComplicatedGeneral}. The Pfaffian is hence determined by the polynomial
\begin{align}
\delta= \tilde{f}^{(1,1)}_{1,1}(y)\tilde{\tilde{f}}^{(1,1)}_{1,1}(y)f^{(2,2)}_{0,1}(y)\,,
\end{align}
with $y=(\vec{z}_2,\vec{z}_3)$ and 
\begin{align}
\tilde{f}^{(1,1)}_{1,1}(y)=\sum_{i=1}^{10}\beta^{(1,1)}_i m_{1,1}^i(y)\,,\quad
\tilde{\tilde{f}}^{(1,1)}_{1,1}(y)=\sum_{i=1}^{10}\tilde{\beta}^{(1,1)}_i m_{1,1}^i(y)\,,\quad
f^{(2,2)}_{0,1}(y)=\sum_{i=1}^{5}\beta^{(2,2)}_i m_{0,1}^i(y)\,.
\end{align}
Next, we find the 32 curves in the curve class of the first $\bP^1$, insert their position into $\delta$ and consider the sum
\begin{align}
\sum_{i=1}^{32}\lambda_i\delta_{i}\,.
\end{align}
Note that $\delta_{i}$ contains 500 terms trilinear in the moduli $\beta$. We find that out of these 500 terms, only 31 are linearly independent. Hence, there does exist an assignment for the $\lambda_i$ such that all contributions cancel.

\subsubsection{The GLSM computation}
The GLSM model for this example is summarized in Table~\ref{tab:GLSMChargesVanishingExamples2}.
\begin{table}[t]
\centering
\begin{tabular}{|c||c@{~~}c@{~~}c@{~~}|c@{~~}c@{~~}c@{~~}||c@{~~}c@{~~}c@{~~}c@{~~}c@{~~}|c@{~~}c@{~~}||c@{~~}|c@{~~}|}
\hline
&$\cZ_{1,2}$ & $\cZ_{3,4}$ & $\cZ_{5,6,7,8,9}$ & $P_1$  & $P_2$ & $P_3$ &$A_1$ & $A_2$ & $A_3$ & $A_4$ & $A_5$ & $B_1$&$B_2$&$S$&$\Xi$\\
\hline
$\U1_1$ & 1&0&0  &-1&-1&0  	& 2&1&0&0&0		&-2&-1		&1 		& -1	\\
$\U1_2$ & 0&1&0  &-2&0&0		& 1&0&1&0&0		&-2&0  		&0		&0	\\
$\U1_3$ & 0&0&1  &-1&-2&-2	& 0&2&0&1&1		&-1&-3  		&-1		&1	\\
\hline
\end{tabular}\newline
\caption{GLSM charges of the monad bundle on CICY 7555.}
\label{tab:GLSMChargesVanishingExamples2}
\end{table}
We find that all GLSM anomalies cancel, so this example has an anomaly-free GLSM description.
From Table~\ref{tab:GLSMChargesVanishingExamples2} we note that the spectator field $S$ has charge $+1$ relative to the first U(1) factor. 
Hence, it satisfies the non-compactness criteria in Eq.~\eqref{eq:BP_Conditions2} and has a zero mode on the first $\bP^1$, 
which decompactifies the instanton moduli space. This means that the general vanishing results of~\cite{Silverstein:1995re, Basu:2003bq, Beasley:2003fx}
cannot be directly applied here. Nevertheless, our geometric analysis has shown that a cancellation of the 32 contributions from the various curves in the first curve class is possible.


\section{Conclusion}
\label{sec:Conclusions}
In this paper, we have considered two main tasks which arise in the context of heterotic instanton superpotential calculations. Firstly, we have developed geometric methods to calculate instanton superpotentials for different realizations of the bundle and for cases with a complicated structure of the bundle moduli space. Secondly, we have analyzed how the results of such geometric calculations relate to the vanishing theorems of Refs.~\cite{Silverstein:1995re, Basu:2003bq, Beasley:2003fx}.

We have shown how to compute Pfaffians, as a function of bundle moduli, for different bundle constructions, specifically for extension bundles, double extensions bundles and for monads. A technical difficulty is the explicit description of the bundle moduli. While the aforementioned bundle constructions involve obvious classes of polynomials whose coefficients can serve as bundle moduli these are usually an over-parametrization. Removing this over-parametrization is essential for a reliable calculation of the Pfaffians as a function of bundle moduli 
and we have shown how to carry this out explicitly. 

If all Pfaffians contributing to a particular second homology class are linearly independent as a function of the bundle moduli, then the corresponding term in the instanton superpotential must be non-zero. Our ability to compute the functional dependence of Pfaffians means that we can explicitly apply this criterion. We have carried this out for a large number of examples, some of which have been described in this paper, with a somewhat surprising outcome. Even though these examples appear to be satisfying the conditions underlying the vanishing theorems~\cite{Silverstein:1995re, Basu:2003bq, Beasley:2003fx}, in many cases the Pfaffians turn out to be linearly independent and, hence, the instanton superpotential must be non-zero.

We have proposed that this apparent contradiction is resolved by taking into account compactness of the instanton moduli space, which is one of the more subtle requirements for the vanishing theorems to be applicable. Currently, the only explicit criterion for compactness of the instanton moduli space is due to Bertolini and Plesser~\cite{Bertolini:2014dna} and is formulated within the framework of GLSMs. Hence, in order to test our proposal, we have studied the relation between geometric models (based on monad bundles) and their GLSM description.

A significant technical difficulty is that the GLSM anomaly conditions are stronger than the geometric ones and that they depend on the specific realization of the geometric model.  For most geometric realizations the associated GLSM is anomalous and it is not clear to us when alternative realizations with a non-anomalous GLSM exist. In practice, we have dealt with this by considering a large number of geometric models and by extracting the relatively small sub-set which does have a non-anomalous GLSM realization.

For this sub-set we have shown that there is no contradiction between the geometric calculation and the vanishing theorems. To summarize this in more details, let us first consider the models within this sub-set with a compact instanton moduli space. For such cases, the instanton superpotential must be zero from the vanishing theorems and this is indeed consistent with the geometric approach, since the Pfaffians always turn out to be linearly dependent. The results for models with a non-compact instanton moduli space are somewhat unexpected. It turns out all these models have linearly dependent Pfaffians as well. For such cases, neither the GLSM nor the geometric approach tell us directly whether the instanton superpotential is zero or non-zero. In particular, the geometric approach fails to make a definite prediction due to the unknown pre-factors $\lambda_i$ of the Pfaffians. However, linear dependence of the Pfaffians is a very non-trivial feature, given their complicated dependence on the bundle moduli. On this basis we have argued that the instanton superpotential for these cases is likely to vanish, a feature which is not explained by the vanishing theorems.

It would be desirable to work out the relation between geometric instanton calculations and vanishing theorems more systematically than we have been able to achieve in this paper. This may well require a better understanding of how geometric models can be converted into anomaly-free GLSMs.


\section*{Acknowledgments}
We would like to thank Chris Beasley, Philip Candelas, Antonella Grassi, Albrecht Klemm, Jock McOrist, Tony Pantev and Balazs Szendroi for helpful discussions. E.I.B.~is supported in part by the ARC Discovery Project DP200101944. B.A.O.~is supported in part by DOE No. DE-SC0007901 and SAS Account 020-0188-2-010202-6603-0338. F.R.~thanks the University of Pennsylvania for hospitality. 

\providecommand{\href}[2]{#2}
\end{document}